\documentclass[11pt,a4paper]{article}
\pdfoutput=1
\usepackage{jheppub}

\usepackage{etoolbox}% http://ctan.org/pkg/etoolbox
\makeatletter
\patchcmd{\maketitle}{\@fpheader}{\vspace{-15mm}}{}{}
\makeatother

\usepackage[normalem]{ulem}
\usepackage{amsmath}
\usepackage{amssymb}
\usepackage{amscd}
\usepackage{enumerate}
\usepackage{amsfonts}
\usepackage{epsfig}
\usepackage{mathtools}
\usepackage{yfonts}
\usepackage{bbold}
\usepackage{breqn}
\usepackage{multirow}

\definecolor{davecolor}{rgb}{0.95,  0.5,  0.2}

\definecolor{darkgreen}{rgb}{0,0.5,0}
\definecolor{darkblue}{rgb}{0,0,0.6}
\definecolor{purple}{rgb}{0.4,0.15,0.21}
\definecolor{black}{rgb}{.2,.2,.2}

\usepackage{dsfont}

\usepackage{graphicx}
\usepackage{bm}
\usepackage{xcolor}

\definecolor{davecolor}{rgb}{0.95,  0.5,  0.2}

\def\({\left(}
\def\){\right)}
\def\[{\left[}
\def\]{\right]}
\def\<{\langle}
\def\>{\rangle}

\newcommand\half{{\ensuremath{\frac{1}{2}}}}
\newcommand\p{\ensuremath{\partial}}

\newcommand{\be}{\begin{equation}}
\newcommand{\ee}{\end{equation}}
\newcommand{\bea}{\begin{eqnarray}}
\newcommand{\eea}{\end{eqnarray}}
\newcommand{\benn}{\begin{equation*}}
\newcommand{\eenn}{\end{equation*}}
\newcommand{\bwt}{\begin{widetext}}
\newcommand{\ewt}{\end{widetext}}

\newcommand{\bi}{\begin{itemize}}
\newcommand{\ei}{\end{itemize}}
\newcommand{\ben}{\begin{enumerate}}
\newcommand{\een}{\end{enumerate}}
\newcommand{\bca}{\begin{cases}}
\newcommand{\eca}{\end{cases}}
\newcommand{\bln}{\begin{align}}
\newcommand{\eln}{\end{align}}
\newcommand{\bst}{\begin{split}}
\newcommand{\est}{\end{split}}
\newcommand{\ba}{\begin{aligned}}
\newcommand{\ea}{\end{aligned}}

\newcommand\Ga{{\ensuremath{{\Gamma}}}}
\newcommand\de{{\ensuremath{{\delta}}}}
\newcommand\De{{\ensuremath{{\Delta}}}}

\newcommand{\sgn}{{\rm sgn}}
\newcommand\G{\Gamma}

\newcommand{\zb}{\bar{z}}
\newcommand{\fr}{\frac}
\newcommand\ra{{\rightarrow}}

\newcommand\bA{{\bf A}}
\newcommand\bB{{\bf B}}
\newcommand\bC{{\bf C}}

\usepackage{empheq}
\usepackage{enumitem} 

\newcommand{\Ical}{{\cal I}}

\newcommand{\Mcal}{{\cal M}}
\newcommand{\Ocal}{{\cal O}}
\newcommand{\Dmax}{\Delta_{\max}}
\newcommand{\hb}{\bar{h}}
\newcommand\kp{k^\prime}
\newcommand\bk{{\bf k}}
\newcommand\bkp{{\bf k}^\prime}
\newcommand\bfm{{\bf m}}
\newcommand\ET{\textrm{ET}}

\title{Momentum space CFT correlators for Hamiltonian truncation}
\author[a]{Nikhil Anand,} 
\author[b,c]{Zuhair U. Khandker,}
\author[d,e]{Matthew T. Walters}
\affiliation[a]{Department of Physics, McGill University, Montr\'{e}al, QC H3A 2T8, Canada} 
\affiliation[b]{Department of Physics, University of Illinois, Urbana, IL 61801, U.S.A.} 
\affiliation[c]{Department of Physics, Boston University, Boston, MA 02215, U.S.A.} 
\affiliation[d]{Theoretical Physics Department, CERN, 1211 Geneva 23, Switzerland}
\affiliation[e]{Institute of Physics, \'{E}cole Polytechnique F\'{e}d\'{e}rale de Lausanne (EPFL), CH-1015 Lausanne, Switzerland \vspace{5mm}}

\abstract{We consider Lorentzian CFT Wightman functions in momentum space. In particular, we derive a set of reference formulas for computing two- and three-point functions, restricting our attention to three-point functions where the middle operator (corresponding to a Hamiltonian density) carries zero spatial momentum, but otherwise allowing operators to have arbitrary spin. A direct application of our formulas is the computation of Hamiltonian matrix elements within the framework of conformal truncation, a recently proposed method for numerically studying strongly-coupled QFTs in real time and infinite volume. Our momentum space formulas take the form of finite sums over ${}_2F_1$ hypergeometric functions, allowing for efficient numerical evaluation. As a concrete application, we work out matrix elements for 3d $\phi^4$-theory, thus providing the seed ingredients for future truncation studies. }

\begin{document}
\maketitle

%%%%%%%%%%%%%%%%%%%%%%%%%%%%%%%%%%%%%%%%%%%%%%%%%%%%%%%%%%%%%%%%%%%%%%%%%%%%%%%%%%%%%%%%%%%%%%%%%%%%
\section{Introduction and summary}
\label{sec:intro}

Accessing real-time dynamics in strongly-coupled quantum field theories (QFTs) is a persistent challenge across many areas of physics. For generic nonperturbative systems, \emph{i.e.}, those lacking a clear expansion parameter or large amounts of symmetry, one must usually rely on numerical methods to make progress. However, even with numerics, accessing truly dynamical quantities (such as time-dependent correlation functions, spectral densities, and quantum wavefunctions of states) is still notoriously difficult. 

\emph{Conformal truncation}~\cite{Katz:2016hxp} is a recently-proposed numerical method for studying strongly-coupled QFTs that is formulated directly in real time and infinite volume to facilitate the computation of dynamics. The basic idea is to view the QFT of interest as the IR limit of a UV conformal field theory (CFT) that has been deformed by one or more relevant operators $\Ocal_R$. At the level of the Hamiltonian,
\be
H_{\textrm{QFT}} = H_{\textrm{CFT}} + \lambda V = H_{\textrm{CFT}} + \lambda \int d^{d-1}\vec{x}\, \Ocal_{R}(\vec{x}),
\label{eq:H}
\ee
where $d$ is the number of spacetime dimensions and $\vec{x}$ denotes spatial directions. Then, one implements a version of \emph{Hamiltonian truncation} starting from the UV CFT that reconstructs the RG flow to access the IR QFT. 

Hamiltonian truncation refers to an array of QFT methods that all share the same basic strategy. First, the QFT Hamiltonian is expressed in a chosen basis. Second, the Hamiltonian is truncated to a finite size according to some prescription. Finally, the truncated Hamiltonian is diagonalized (usually numerically) to obtain an approximation to the physical spectrum and eigenstates of the QFT. For a recent review of these methods, see~\cite{James:2017cpc}. Different Hamiltonian truncation methods differ precisely in their choice for the basis and the prescription used for truncation. 

Conformal truncation is a variant of Hamiltonian truncation where one uses states in the UV CFT to construct a basis. Specifically, the conformal truncation basis consists of Fourier transforms of CFT primary operators,
\be
|\Ocal(P)\rangle \equiv \int d^dx\, e^{-iP\cdot x} \Ocal(x) |0\rangle.
\label{eq:CTstate}
\ee
The Hamiltonian is expressed in this basis, and truncation occurs in the scaling dimension $\Delta$ of the operators $\Ocal(x)$: only operators with $\Delta$ below some cutoff $\Dmax$ are included. The resulting Hamiltonian matrix is diagonalized numerically in order to obtain an approximation to physical observables like the spectrum and correlation functions. Finally, one looks for convergence in computed observables as the threshold $\Dmax$ is increased. For recent implementations of conformal truncation, see~\cite{Katz:2013qua,Katz:2014uoa,Anand:2017yij,Delacretaz:2018xbn}.

A direct consequence of the basis choice in~(\ref{eq:CTstate}) is that inner products and Hamiltonian matrix elements are given by Fourier transforms of CFT two- and three-point functions. Formally, the inner product between two states is given by 
\be
\langle \Ocal(P) | \Ocal^\prime(P^\prime ) \rangle = (2\pi)^d \delta^d(P-P^\prime) \int d^dx \, e^{iP\cdot x} \langle \Ocal(x)\Ocal^\prime(0) \rangle,
\label{eq:Inner}
\ee
while matrix elements of the Hamiltonian deformation $V$ are given by
\be
\langle \Ocal(P) | V | \Ocal^\prime(P^\prime ) \rangle = (2\pi)^{d-1} \delta^{d-1}(\vec{P}-\vec{P}^\prime) \int d^dx \, d^dx^\prime \, e^{i(P\cdot x - P^\prime\cdot x^\prime)} \langle \Ocal(x) \Ocal_R(0) \Ocal^\prime(x^\prime) \rangle. 
\label{eq:ME}
\ee
Thus, at a schematic level, the basic ingredients of conformal truncation are simply CFT correlators in momentum space. However, to fully define (\ref{eq:ME}), we need to specify the quantization scheme.

Conformal truncation uses lightcone quantization~\cite{Dirac:1949cp}. Our conventions for lightcone coordinates are
\begin{equation}
\begin{aligned}
& x^\pm = \frac{1}{\sqrt{2}}\left( x^0 \pm x^1 \right), \hspace{10mm} \vec{x}^\perp = (x^2,\cdots, x^{d-1}), \\
& \hspace{19mm} ds^2 = 2dx^+ dx^- - d\vec{x}^{\perp 2}.
\end{aligned}
\label{eq:metric}
\end{equation}
In lightcone quantization, one considers $x^+$ to be time and $\vec{x} = (x^-,\vec{x}^\perp)$ to be spatial. The lightcone momenta are defined by $P_{\pm} \equiv \left( P_0 \pm P_1 \right) / \sqrt{2}$. In particular, the lightcone Hamiltonian is $P_+$, with 
\be
V = \int d^{d-1} \vec{x} \, \Ocal_R(x^+ =0, \vec{x}). 
\ee
This is the quantization scheme defining (\ref{eq:ME}). Interestingly, as we will see, lightcone coordinates are very natural for evaluating Fourier transforms of CFT correlators because of a factorization that occurs in the $x^\pm$ integrations. This factorization leads to simple expressions in two dimensions and motivates our strategy for higher dimensions.\footnote{In the context of Hamiltonian truncation, lightcone quantization offers several advantages compared to the more familiar equal-time quantization. In particular, all Hamiltonian matrix elements involving the vacuum vanish in lightcone quantization, such that the vacuum is not renormalized~\cite{Leutwyler:1970wn,Maskawa:1975ky,Brodsky:1997de}. This is a consequence of the fact that the lightcone momentum $P_{-}=0$ for the vacuum but is strictly positive for any non-vacuum state, so that the vacuum cannot mix with other states by momentum conservation. For free CFTs, this means that matrix elements corresponding to particle creation from vacuum must vanish, which is a significant simplification compared to equal-time quantization. The caveat to all of these statements is the possible effects of `zero modes'. There is now a better understanding of how to systematically account for zero modes with an effective Hamiltonian~\cite{Fitzpatrick:2018ttk,Burkardt} and how to match results between lightcone and equal-time quantization~\cite{Fitzpatrick:2018xlz,Chabysheva:2016wvl,Chabysheva:2018wxr}.}  

The correlators appearing on the right-hand sides of (\ref{eq:Inner})-(\ref{eq:ME}) are \emph{Wightman functions}, with operators ordered as written. This fixed ordering of the correlator is crucial for constructing well-defined in- and out-states for the Hamiltonian matrix elements and is obtained by using a particular $i\epsilon$ prescription when performing the Fourier transform integrals to momentum space, which we will review.

Any conformal truncation effort will rely on our ability to efficiently compute inner products and Hamiltonian matrix elements. Schematically, (\ref{eq:Inner})-(\ref{eq:ME}) will take the form 
\be
\ba
\langle \Ocal(P) | \Ocal^\prime(P^\prime ) \rangle &= (2\pi)^d \delta^d(P-P^\prime) \, \Ical_{\Ocal\Ocal^\prime}(P,P^\prime), \\[5pt]
\<\Ocal(P)|V|\Ocal'(P')\> &= (2\pi)^{d-1} \de^{d-1}(\vec{P}-\vec{P}') \, C_{\Ocal\Ocal'\Ocal_R} \, \Mcal^{\Ocal_R}_{\Ocal\Ocal'}(P,P'),
\ea
\label{eq:Schematic}
\ee
where $C_{\Ocal\Ocal'\Ocal_R}$ are theory-dependent OPE coefficients, while $\Ical_{\Ocal\Ocal^\prime}$ and $\Mcal^{\Ocal_R}_{\Ocal\Ocal'}$ are kinematic functions of momenta, whose structure is completely fixed by conformal symmetry, based on the scaling dimensions and spins of $\Ocal$, $\Ocal'$, and $\Ocal_R$. 

The goal of this work is to derive reference formulas for $\Ical_{\Ocal\Ocal^\prime}$ and $\Mcal^{\Ocal_R}_{\Ocal\Ocal'}$ in 1+1 and 2+1 dimensions. Then, as a direct application, we will use the reference formulas to compute inner products and matrix elements for 3d $\phi^4$-theory, thus providing the seed ingredients for future truncation studies of this theory. 

From (\ref{eq:Inner})-(\ref{eq:ME}), we see that the kinematic function $\Ical_{\Ocal\Ocal^\prime}$ is a general CFT two-point Wightman function in momentum space, while $\Mcal^{\Ocal_R}_{\Ocal\Ocal'}$ corresponds to a three-point function, with the restriction that the middle operator has fixed $x^+$ and zero spatial momentum, because it corresponds to a Hamiltonian density. This restriction on the spatial momentum is particularly constraining in lightcone quantization, because it forces the momentum of the middle operator to be \emph{null},
\be
P^2 = 2P_+ P_- - |\vec{P}_\perp|^2 \, \Rightarrow \, 0 \quad \textrm{when} \quad P_-,\vec{P}_\perp = 0.
\ee
Computing Hamiltonian matrix elements in lightcone quantization is thus equivalent to computing CFT three-point Wightman functions in momentum space where the middle operator has null momentum.

This restriction on the momentum greatly simplifies the task of Fourier transforming, as well as the resulting momentum space expressions. However, we must allow for the external operators $\Ocal$ and $\Ocal'$ to have \emph{arbitrary spin}, since they are general operators in a CFT. This makes things more challenging in higher dimensions.
 
In 2d, we directly compute the inner products and matrix elements of primary operators with general spin. In 3d, we compute Fourier transforms of a simple but complete basis of tensor structures that can appear in correlation functions of arbitrary-spin primaries. In other words, we compute contributions to $\Ical_{\Ocal\Ocal^\prime}$ and $\Mcal^{\Ocal_R}_{\Ocal\Ocal'}$ coming from a basis of tensor structures that can appear in $\< \Ocal \Ocal^\prime \>$ and $\< \Ocal \Ocal_R \Ocal^\prime \>$. Taking appropriate linear combinations of our results, one can obtain the inner products and matrix elements of general primary operators. Next, we apply our 3d formulas to $\phi^4$-theory and compute inner products and matrix elements of so-called ``monomial'' operators, which are products of derivatives acting on multiple insertions of $\phi$. Again, one can take linear combinations of these results to obtain expressions for general primaries.  

The main results of this paper are summarized as follows:
\begin{table}[h!]
\centering
\begin{tabular}{|c|l|}
\hline
\multirow{2}{*}{$d=1+1$ (Primary operators):} & -- Two-point functions \eqref{eq:2dInner} \\ 
                  & -- Three-point functions \eqref{eq:2dME} \\ \hline
$d=2+1$ (Complete basis of & -- Two-point functions \eqref{eq:IFinal} \\  
                  tensor structures): & -- Three-point functions \eqref{eq:JFinal} \\ \hline
\multirow{3}{*}{3d $\phi^4$-theory:} & -- Inner products \eqref{eq:InnerFinal} \\  
                  & -- Matrix elements for $\phi^2$ \eqref{eq:MassFinal} \\  
                  & -- Matrix elements for $\phi^4$ \eqref{eq:NNFinal}, \eqref{eq:22Final}, \eqref{eq:NN2Final}, \eqref{eq:13Final} \\ \hline
\end{tabular}
\end{table}

\noindent These formulas are meant to serve as a reference for future truncation applications.

Our basic strategy in 3d is to perform the Fourier integral over the perpendicular direction $x^\perp$ first, thus reducing the problem back down to 2d. We find that conformal truncation matrix elements in 3d are effectively sums over 2d matrix elements. More concretely, 2d matrix elements are expressible in terms of the $\phantom{}_2F_1$ hypergeometric function, see~(\ref{eq:2dME}). In 3d, while the reference formula for Fourier transforms in (\ref{eq:JFinal}) looks complicated, note that it is a closed-form expression that can be applied to operators with arbitrary spin, and is ultimately just a sum of $\phantom{}_2F_1$'s.  

For $\phi^4$-theory, the UV CFT is free massless scalar field theory, so one could alternatively use Fock space methods to compute conformal truncation matrix elements. We have checked that the $\phi^4$-theory matrix elements presented here agree with results obtained using independent Fock space techniques, providing a highly nontrivial check of our formulas. The Fock space methods are interesting in their own right and require their own analytical machinery, which we will cover in a separate publication. However, it is worth emphasizing that the momentum space formulas presented in this work are much more efficient than Fock space methods for computing matrix elements. Moreover, in other applications where the UV CFT is not free, Fock space methods are not applicable, and working directly with momentum space correlators provides the only path forward. 

Recently, there has been much interest in studying CFT correlators in momentum space. Indeed, momentum space is the natural arena for many types of observables. Example contexts include cosmology~\cite{Maldacena:2011nz,Creminelli:2012ed,Mata:2012bx,Kundu:2014gxa,Arkani-Hamed:2015bza,Kundu:2015xta,Shukla:2016bnu,Arkani-Hamed:2018kmz,Sleight:2019mgd,Sleight:2019hfp,Baumann:2019oyu}, `conformal collider' physics~\cite{Hofman:2008ar,Belitsky:2013xxa,Belitsky:2013bja,Belitsky:2013ofa,Kologlu:2019mfz}, anomalies~\cite{Gillioz:2016jnn,Coriano:2017mux,Gillioz:2018kwh,Coriano:2018zdo}, and quantum critical transport~\cite{Chowdhury:2012km,Huh:2013vga,Jacobs:2015fiv,Lucas:2016fju,Myers:2016wsu,Lucas:2017dqa}. Momentum space is also useful for computing conformal blocks and the conformal bootstrap~\cite{Isono:2018rrb,Gillioz:2018mto,Erramilli:2019njx,Isono:2019wex}. In the context of conformal truncation, momentum space CFT correlators are the gateway for accessing general strongly-coupled QFTs. Indeed, the Hamiltonian matrix elements in~(\ref{eq:ME}) encode information about renormalization group flows emanating from the CFT. The more of these matrix elements we can compute, the more power we have to extract that information and make predictions beyond the CFT.  

Interestingly, though, CFT correlators in \emph{Lorentzian} momentum space, which is our interest here, are much less understood than their Euclidean counterparts, which have been studied extensively (\emph{e.g.}, \cite{Coriano:2013jba,Bzowski:2013sza,Bzowski:2015pba,Bzowski:2017poo,Bzowski:2018fql,Coriano:2018bsy,Isono:2019ihz,Bzowski:2019kwd}). In principle, one expects to be able to obtain Lorentzian correlators from Euclidean ones via Wick rotation. However, doing this in practice is subtle for momentum space Wightman functions, and results so far have been restricted to operators with low spin~\cite{Chowdhury:2012km,Bautista:2019qxj,Gillioz:2019lgs}. It is worth reiterating that in this paper, we are studying the structure of Lorentzian CFT correlators in momentum space, but with the restriction that the middle operator in three-point functions has zero spatial momentum. Thus (\ref{eq:ME}) is a special case of the most general Fourier transform of a CFT three-point function. It is intriguing that in our scenario, we are able to do computations for arbitrary spin. It would be interesting to see if the strategies used here can be combined with those of~\cite{Chowdhury:2012km,Bautista:2019qxj,Gillioz:2019lgs} to tackle the most general momentum space correlators for any spins. 

As a final remark, we mention that working in Lorentzian spacetime requires keeping careful track of any and all phases arising from Wick rotation. They are important, and affect the final formulas. We have made every effort to be precise with phases so that $i$'s and minus signs in every formula can be trusted.\footnote{We welcome attention brought to any errors in this regard.}

This paper is organized as follows. In section~\ref{sec:epsilon}, we review the $i\epsilon$ prescription needed for constructing Wightman functions. In section~\ref{sec:2Dformulas}, we present our 2d momentum space formulas. This is a useful warm-up before deriving our 3d formulas in section~\ref{sec:3Dformulas}. Our approach in higher dimensions, while simple, is admittedly brute-force and leaves something to be desired. In section~\ref{sec:AdS}, we briefly discuss an alternative approach, using AdS Witten diagrams to compute momentum space correlators in general dimension $d$ for the specific case of scalar operators. If this approach can be generalized to spinning operators, it may provide an alternative strategy. In section~\ref{sec:FFT}, we use our momentum space formulas to compute matrix elements for 3d $\phi^4$-theory. These are the seed ingredients for upcoming numerical studies of this theory. We conclude in section~\ref{sec:conclusion} with an outlook for the future. Appendix~\ref{sec:notation} contains a summary of notation introduced in the body of the paper, for the reader's convenience, while appendix~\ref{sec:useful} contains some useful formulas for Lorentzian AdS propagators.

%%%%%%%%%%%%%%%%%%%%%%%%%%%%%%%%%%%%%%%%%%%%%%%%%%%%%%%%%%%%%%%%%%%%%%%%%%%%%%%%%%%%%%%%%%%%%%%%%%%%
\section{$i\epsilon$ prescription for Wightman functions}
\label{sec:epsilon}

The inner product and matrix elements of our conformal truncation basis states are specifically defined as the Fourier transforms of Wightman functions, with a particular fixed ordering for the operators. In general, a Lorentzian correlator can be defined as the analytic continuation of a Euclidean correlation function to imaginary Euclidean time,
\be
\tau_i  \to i t_i,
\ee
where $t_i$ is the real Lorentzian time associated with the local operator $\Ocal_i$.

To obtain a particular ordering of operators in the resulting Lorentzian correlator, we introduce an infinitesimal real Euclidean time $\epsilon_i$ for each operator and define
\be
\<\Ocal_1(t_1,\vec{x}_1) \cdots \Ocal_n(t_n,\vec{x}_n)\> \equiv \lim_{\epsilon_i \to 0} \<\Ocal_1(i(t_1 - i\epsilon_1),\vec{x}_1) \cdots \Ocal_n(i(t_n - i\epsilon_n),\vec{x}_n)\>,
\ee
where the limit is taken with $\epsilon_1 > \cdots > \epsilon_n$ (\emph{e.g.},~\cite{Haag:1992hx,Hartman:2015lfa}). In this definition, the expression on the right-hand side is an analytically continued Euclidean correlator, and the expression on the left-hand side is the resulting Lorentzian Wightman function.

For example, let's consider the two-point function of some scalar primary operator in general dimension $d$. The original Euclidean correlator takes the simple form,
\be
\<\Ocal(\tau,\vec{x}) \Ocal(0)\> = \fr{1}{(\tau^2 + |\vec{x}|^2)^\De}.
\ee
To obtain a Lorentzian correlator where the operator acting at the origin is on the right (\textit{i.e.}~acts first), we need to consider the analytic continuation,
\be
\tau \to i(t - i\epsilon),
\ee
with positive $\epsilon$, resulting in the Lorentzian correlator
\be
\<\Ocal(t,\vec{x})\Ocal(0)\> = \fr{e^{-i\pi \De}}{(t^2 - |\vec{x}|^2 - i\epsilon \, \sgn t)^\De}.
\label{eq:2ptWightman}
\ee

In this work, we will typically be working in lightcone coordinates, in which case it is more convenient to write the original Euclidean correlator in terms of the holomorphic coordinates
\be
z = \fr{1}{\sqrt{2}}(\tau + ix^1), \quad \zb = \fr{1}{\sqrt{2}}(\tau - ix^1).
\ee
which, for this two-point function example, results in the Euclidean expression
\be
\<\Ocal(z,\zb,\vec{x}^\perp) \Ocal(0)\> = \fr{1}{(2z\zb + |\vec{x}^\perp|^2)^\De}.
\ee
To obtain the same Wightman function as above (with $\Ocal(0)$ on the right), we therefore need to perform the analytic continuation
\be
z \to i(x^+ - i\epsilon), \quad \zb \to i(x^- - i\epsilon),
\ee
resulting in
\be
\boxed{\<\Ocal(x)\Ocal(0)\> = \fr{e^{-i\pi\De}}{\Big(2(x^+ - i\epsilon)(x^--i\epsilon) - |\vec{x}^\perp|^2\Big)^\De},}
\ee
which is equivalent to eq.~\eqref{eq:2ptWightman}.

Similarly, we can consider a scalar three-point function, with the original Euclidean correlator
\be
\<\Ocal_1(x_1) \Ocal_2(x_2) \Ocal_3(x_3)\> = \fr{C_{123}}{x_{12}^{\De_1+\De_2-\De_3} x_{23}^{\De_2+\De_3-\De_1} x_{13}^{\De_1+\De_3-\De_2}}.
\ee
If we want to construct the corresponding Wightman function where $\Ocal_1$ is on the far left and $\Ocal_3$ is on the far right, then we need to perform the analytic continuation
\be
\tau_1 \to i(t_1-i\epsilon), \quad \tau_2 \to it_2, \quad \tau_3 \to i(t_3 + i\epsilon).
\ee
For this particular ordering, the resulting Lorentzian correlator is
\be
\boxed{
\ba
&\<\Ocal_1(x_1) \Ocal_2(x_2) \Ocal_3(x_3)\> \\
& \quad = \fr{C_{123} \, e^{-\fr{i\pi}{2}(\De_1+\De_2+\De_3)}}{\Big(2(x_{12}^+ - i\epsilon)(x_{12}^--i\epsilon) - |\vec{x}_{12}^\perp|^2\Big)^{\fr{\De_1+\De_2-\De_3}{2}}  \Big(2(x_{23}^+ - i\epsilon)(x_{23}^--i\epsilon) - |\vec{x}_{23}^\perp|^2\Big)^{\fr{\De_2+\De_3-\De_1}{2}}} \\
& \quad \quad \times \fr{1}{\Big(2(x_{13}^+ - i\epsilon)(x_{13}^--i\epsilon) - |\vec{x}_{13}^\perp|^2\Big)^{\fr{\De_1+\De_3-\De_2}{2}}}.
\ea
}
\ee

%%%%%%%%%%%%%%%%%%%%%%%%%%%%%%%%%%%%%%%%%%%%%%%%%%%%%%%%%%%%%%%%%%%%%%%%%%%%%%%%%%%%%%%%%%%%%%%%%%%%
\section{2d momentum space formulas}
\label{sec:2Dformulas}

In this section, we evaluate the Fourier transforms \eqref{eq:Inner} and \eqref{eq:ME} in $d=2$ spacetime dimensions. As we will see, the results in 2d contain all of the essential features we will need to generalize to higher dimensions.

\subsection{Two-point functions}
\label{subsec:2D2PF}

In 2d, the Lorentzian CFT two-point function for a general primary operator $\Ocal$ is
\be
\langle \Ocal(x)\Ocal(0) \rangle = \frac{e^{-i\pi\Delta}}{(x^+-i\epsilon)^{2h}(x^--i\epsilon)^{2\hb}},
\ee
where $h,\hb$ are the conformal weights of $\Ocal$, which are related to the scaling dimension $\De$ and spin $J$ via
\be
\De = h+\hb, \quad J = |h-\hb|.
\ee
Note that we are using the $i\epsilon$ prescription for Wightman correlators explained in the previous section. Since the correlator factorizes in $x^\pm$, the Fourier transform in lightcone coordinates likewise factorizes into a product of two independent integrals, 
\be
\int d^2x \, e^{iP\cdot x} \<\Ocal(x) \Ocal(0)\> = \int d^2x \, e^{iP\cdot x} \frac{e^{-i\pi\De}}{(x^+-i\epsilon)^{2h}(x^--i\epsilon)^{2\hb}} = I_h(P_+) \, I_{\hb}(P_-),
\ee
where we have defined the general one-dimensional integral
\be
I_h(P) \equiv  e^{-i\pi h} \int_{-\infty}^\infty dx \, \frac{e^{iPx}}{(x-i\epsilon)^{2h}}.
\label{eq:IhPDef}
\ee

\begin{figure}[t!]
\begin{center}
\includegraphics[width=0.5\textwidth]{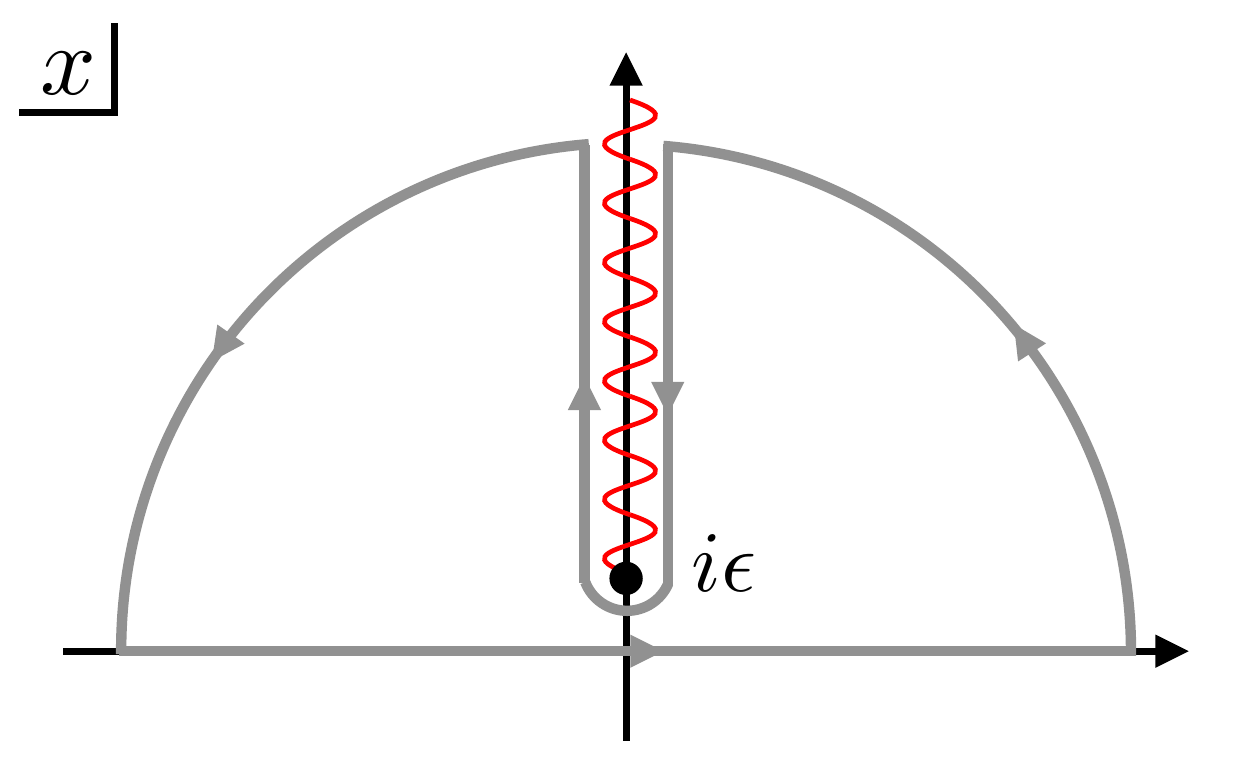}
\caption{Integration contour for evaluating eq.~\eqref{eq:IhPDef}. The $i\epsilon$ prescription for this Wightman function places the branch point in the upper half of the complex plane, which ensures that the Fourier transform only has support for physical lightcone momentum $P>0$. The discontinuity along this branch cut gives the resulting momentum space expression in eq.~\eqref{eq:IhP}.}
\label{fig:TwoPtContour} 
\end{center}
\end{figure}

As shown in figure~\ref{fig:TwoPtContour}, the integrand for $I_h(P)$ contains a branch point at $x=i\epsilon$, since generically $2h$ is not an integer. If $P < 0$, we can close the contour in the lower half of the complex plane, which contains no singularities, such that the integral vanishes. The $i\epsilon$ prescription for this Wightman function therefore gives rise to a Heaviside step function $\Theta(P)$, which ensures that the Fourier transform only has support for positive lightcone momentum $P$.

For $P > 0$, we can evaluate $I_h(P)$ by integrating along the contour shown in figure~\ref{fig:TwoPtContour}, resulting in the expression
\be
I_h(P) = \frac{2\pi P^{2h-1}}{\Gamma(2h)} \, \Theta(P).
\label{eq:IhP}
\ee
Plugging into (\ref{eq:Inner}), we obtain the 2d inner product
\be
\boxed{\<\Ocal(P) | \Ocal (P') \> = (2\pi)^2 \delta^2(P-P) \cdot \frac{4\pi^2 P_+^{2h-1} P_-^{2\hb-1}}{\Gamma(2h)\Gamma(2\hb)} \Theta(P_+) \Theta(P_-).}
\label{eq:2dInner}
\ee

\subsection{Three-point functions}
\label{subsec:2D3PF}

Let us now turn to the Hamiltonian matrix elements (\ref{eq:ME}) for $d=2$. The Lorentzian CFT three-point function is
\be
\ba
\langle \Ocal(x) \Ocal_R(0) \Ocal^\prime(x^\prime) \rangle &= C_{\Ocal\Ocal'\Ocal_R} \, e^{-\frac{i\pi}{2}(\Delta+\Delta_R+\Delta^\prime)} \frac{1}{(x^+-i\epsilon)^{A} (-x^{\prime+}-i\epsilon)^B (x^+-x^{\prime +}-i\epsilon)^{C} } \\
& \qquad \qquad \times \frac{1}{(x^- -i\epsilon)^{\bar{A}} (-x^{\prime-}-i\epsilon)^{\bar{B}} (x^--x^{\prime -}-i\epsilon)^{\bar{C}} },
\ea
\ee
where
\be
\ba
A= h+h_R-h^\prime, \hspace{10mm} B = h_R+h^\prime-h, \hspace{10mm} C=h+h^\prime - h_R, \\
\bar{A}= \hb+\hb_R-\hb^\prime, \hspace{10mm} \bar{B} = \hb_R+\hb^\prime-\hb, \hspace{10mm} \bar{C}=\hb + \hb^\prime - \hb_R.
\ea
\ee
Just as for inner products, the expression for 2d matrix elements factorizes into two independent integrals,
\bea
\langle \Ocal(P) | V | \Ocal'(P') \rangle = (2\pi) \delta(P_- - P'_-) \, C_{\Ocal\Ocal'\Ocal_R} \, M_{ABC}(P_+,P_+^\prime) \, M_{\bar{A}\bar{B}\bar{C}}(P_-,P_-^\prime),
\label{eq:2DME1}
\eea
where we have defined
\be
M_{ABC}(P,P^\prime) \equiv e^{-\fr{i\pi}{2}(A+B+C)} \int dx \, dx^\prime \frac{e^{i(Px-P^\prime x^\prime)}}{(x-i\epsilon)^A   (-x^\prime - i\epsilon)^B (x-x^\prime-i\epsilon)^C}.
\ee

Without loss of generality, in evaluating this integral we can assume $P < P^\prime$. Given this assumption, it is convenient to start with the $x^\prime$ integration. As we can see, there are two branch points, at $x' = -i\epsilon$ and $x' = x - i\epsilon$, which are associated with $\Ocal'$ crossing the lightcones of $\Ocal_R$ and $\Ocal$, respectively. The $i\epsilon$ prescription pushes both of these branch points into the lower half of the complex plane, which means that the integral vanishes for $P' < 0$, similar to the two-point function.

For $P'>0$, we can evaluate the $x'$ integral with a similar contour to figure~\ref{fig:TwoPtContour}, though now with discontinuities along two branch cuts. However, a somewhat simpler approach is to first assume the exponents $B$ and $C$ are integers, in which case the singularities simply become poles, evaluate the associated residues, then analytically continue the resulting expression. Following this approach, we obtain the sum
\be
\ba
&\int dx^\prime \frac{e^{-iP^\prime x^\prime}}{ (x^\prime + i\epsilon)^B (x-x^\prime-i\epsilon)^C} \\
&= -2\pi i \, \Theta(P^\prime) \left( \sum_{m=0}^{B-1} \frac{(C)_m (-iP^\prime)^{B-1-m}}{m! \Gamma(B-m) (x-i\epsilon)^{C+m}} - \sum_{m=0}^{C-1} \frac{(B)_m (iP^\prime)^{C-1-m} e^{-iP^\prime x}}{m! \Gamma(C-m) (x-i\epsilon)^{B+m}} \right).
\ea
\ee
Turning to the remaining $x$ integration, we encounter two types of integrals:
\bea
\int dx \frac{e^{iPx}}{(x-i\epsilon)^{A+C+m}} &=& \frac{2\pi e^{i\frac{\pi}{2}(A+C+m)} P^{A+C+m-1}}{\Gamma(A+C+m)} \Theta(P), \\[5pt]
\int dx \frac{e^{i(P-P^\prime)x}}{(x-i\epsilon)^{A+B+m}} &=& 0, \hspace{10mm}  (P < P^\prime). \label{eq:Int2}
\eea
Both of these integrals are of the form of eq.~\eqref{eq:IhPDef}. The first simply has support for physical momentum $P > 0$, but the second integral vanishes under our assumption that $P < P'$, due to the $i\epsilon$ prescription. Putting these pieces together, we obtain the overall expression
\be
\ba
M_{ABC}(P,P^\prime) &= \frac{4\pi^2 P^{A+C-1} P^{\prime B-1}}{\Gamma(A+C)\Gamma(B)} \Theta(P) \Theta(P') \sum_{m=0}^{B-1} (-1)^m {B-1\choose m} \frac{(C)_m}{(A+C)_m} \left( \frac{P}{P^\prime}  \right)^m \\
&= \frac{4\pi^2 P^{A+C-1} P^{\prime B-1}}{\Gamma(A+C)\Gamma(B)} \, \Theta(P) \, \Theta(P') \, {}_2F_1\left( C, 1-B, A+C, {\textstyle\frac{P}{P^\prime}} \right) \qquad (P < P').
\ea
\label{eq:M2d}
\ee

Plugging this back into (\ref{eq:2DME1}), we arrive at the formula for 2d matrix elements,
\be
\ba
&\langle \Ocal(P) | V | \Ocal^\prime(P^\prime ) \rangle \\
&\quad = (2\pi)\delta(P_- - P_-^\prime) \frac{16\pi^4 C_{\Ocal\Ocal^\prime\Ocal_R} P_+^{2h-1}P_-^{2\hb-1}P_+^{\prime h_R+h^\prime-h-1}P_-^{\prime \hb_R+\hb^\prime-\hb-1}}{\Gamma(2h)\Gamma(2\hb)\Gamma(h^\prime+h_R-h)\Gamma(\hb^\prime+\hb_R-\hb)} \\
& \qquad \times \Theta(P_-) \Theta(P^\prime_-) {}_2F_1\left(\hb+\hb^\prime-\hb_R, 1-\hb_R-\hb^\prime+\hb, 2\hb, {\textstyle\frac{P_-}{P_-^\prime}} \right) \\
& \qquad \times \Theta(P_+) \Theta(P^\prime_+) {}_2F_1\left(h+h^\prime-h_R, 1-h_R-h^\prime+h, 2h, {\textstyle\frac{P_+}{P_+^\prime}} \right) \qquad (P_+ < P_+^\prime).
\ea
\label{eq:2d3ptGeneral}
\ee

Note that the overall delta function fixes $P_-=P^\prime_-$, which sets the argument of the first hypergeometric function to $1$.\footnote{The assumption $P < P^\prime$ in (\ref{eq:M2d}) implies that the limit $P_-/P^\prime_- \to 1$ in (\ref{eq:2d3ptGeneral}) is taken from below.} The behavior of the hypergeometric function in this limit is set by the dimension of the middle operator $\Ocal_R$. For $\hb_R > \half$, the function is finite, and collapses into a ratio of gamma functions, resulting in the Hamiltonian matrix element
\be
\boxed{
\ba
&\langle \Ocal(P) | V | \Ocal^\prime(P^\prime ) \rangle \\
&= (2\pi)\delta(P_- - P_-^\prime) \frac{16\pi^4 C_{\Ocal\Ocal^\prime\Ocal_R} \Ga(2\hb_R-1) P_+^{2h-1}P_+^{\prime h_R+h^\prime-h-1}P_-^{\hb+\hb_R+\hb^\prime-2}}{\Gamma(2h)\Gamma(h^\prime+h_R-h)\Gamma(\hb^\prime+\hb_R-\hb)\Ga(\hb+\hb_R-\hb^\prime)\Ga(\hb+\hb_R+\hb^\prime-1)} \\
& \, \times \Theta(P_-) \Theta(P_+) \Theta(P^\prime_+) {}_2F_1\left(h+h^\prime-h_R, 1-h_R-h^\prime+h; 2h; {\textstyle\frac{P_+}{P_+^\prime}} \right) \qquad (P_+ < P_+^\prime,\hb_R > {\textstyle\frac{1}{2}}). 
\ea
}
\label{eq:2dME}
\ee

For $\hb_R \leq \half$, the hypergeometric function diverges in this limit. This divergence arises due to eq.~\eqref{eq:Int2}, which for $P_- = P'_-$ no longer necessarily vanishes, and in fact can be infinite if the exponent in the denominator is too small. This divergence can be regulated by inserting some nonzero momentum $Q_-$ for the Hamiltonian, in which case we recover eq.~\eqref{eq:2d3ptGeneral}, with the overall delta function modified to $\delta(P_- + Q_- - P^\prime_-)$.

This behavior is quite general, and persists in higher dimensions, where operators with dimension $\De_R \leq \fr{d}{2}$ give rise to divergent Hamiltonian matrix elements. In the context of Hamiltonian truncation, this issue is discussed in more detail in \cite{Katz:2016hxp} for the case of free field theory, where these divergences can be eliminated by restricting to a ``Dirichlet'' subspace created by particular linear combinations of primary operators. For the rest of this work, we will mostly ignore this subtlety and focus on the case where $\De_R > \fr{d}{2}$.

%%%%%%%%%%%%%%%%%%%%%%%%%%%%%%%%%%%%%%%%%%%%%%%%%%%%%%%%%%%%%%%%%%%%%%%%%%%%%%%%%%%%%%%%%%%%%%%%%%%%
\section{3d momentum space formulas}
\label{sec:3Dformulas}

We now consider the Fourier transforms (\ref{eq:Inner})-(\ref{eq:ME}) in three spacetime dimensions, with coordinates $\left( x^+, x^-, x^\perp \right)$. Our general strategy will be to perform all of the $x^\perp$ integrals first, thus reducing the 3d expressions to 2d ones, which can subsequently be evaluated using the formulas in the previous section. 

An important simplification we will make throughout is to work in the frame
\be
P_\perp = 0.
\ee
All of the formulas we derive will be in this frame. The expressions for more general reference frames can easily be obtained by applying Lorentz transformations, specifically spatial rotations, to our results. Of course, for truncation applications rotating to a general frame is usually unnecessary. One can always work in the $P_\perp=0$ frame, in which case the formulas below directly apply.

As mentioned in section~\ref{sec:intro}, we will not be explicitly computing the two- and three-point functions of primary operators, as we did for 2d. Instead, we will be computing the Fourier transforms of \emph{monomial} tensor structures with fixed powers of $x^+$, $x^-$, and $x^\perp$. These individual terms can then be combined to reconstruct the correlation functions corresponding to primary operators.

\subsection{Two-point functions}
\label{subsec:3D2PF}

In 3d, CFT Euclidean two-point functions are linear combinations of terms of the general form\footnote{Note that in this section lowercase variables ($a,b,\ldots$) will exclusively be used for exponents which are \emph{integers}, while capitalized variables ($A,B,\ldots$) will be used for more general exponents.}
\be 
\langle \Ocal(z,\zb,x^\perp)\Ocal(0) \rangle \supseteq \frac{z^{a_+}\zb^{a_-}(x^\perp)^{a_\perp}}{(2z\zb + x^{\perp2})^A}.
\ee
If we analytically continue to Lorentzian signature, the corresponding terms are
\be 
\langle \Ocal(x)\Ocal(0) \rangle \supseteq \frac{e^{-\fr{i\pi}{2}(2A - a_+ - a_-)} (x^+-i\epsilon)^{a_+}(x^- - i\epsilon)^{a_-}(x^\perp)^{a_\perp}}{\Big(2(x^+ - i\epsilon)(x^- - i\epsilon) - x^{\perp2} \Big)^A}.
\label{eq:3d2ptWightman}
\ee
In order to obtain the contribution of such a term to the momentum space Wightman function, we need to compute the function
\be
\Ical_\bA(P) \equiv e^{-\fr{i\pi}{2}(2A - a_+ - a_-)} \int d^3x \, e^{iP\cdot x}\, \frac{(x^+)^{a_+}(x^-)^{a_-}(x^\perp)^{a_\perp}}{(x^2)^A},
\ee
where for simplicity we have suppressed the factors of $i\epsilon$ and introduced the shorthand notation
\be
\bA \equiv (a_+,a_-,a_\perp,A).
\label{eq:bA}
\ee
Inner products of (general-spin) primary operators will be linear combinations of the functions $\Ical_\bA(P)$. Specifically, for some coefficients $T_{\Ocal\Ocal^\prime}^{\bA} $,
\be
\langle \Ocal(P) | \Ocal^\prime(P^\prime ) \rangle = (2\pi)^3 \delta^3(P-P^\prime) \sum_{\bA} T_{\Ocal\Ocal^\prime}^{\bA} \, \Ical_\bA(P).
\ee

The first step in evaluating $\Ical_\bA$ is to perform the integral over $x^\perp$. Since we are working in the frame $P_\perp = 0$, the formula we need is
\be
\int_{-\infty}^{\infty} dx \frac{x^n}{(x^2 + \Omega)^\Delta} =  \Big(1+(-1)^n\Big) \frac{\Gamma(\frac{n+1}{2}) \Gamma(\Delta-\frac{n+1}{2})}{2\Gamma(\Delta)\, \Omega^{\Delta-\frac{n+1}{2}}}. 
\label{eq:PerpIntegral}
\ee
In this frame, the Fourier transform vanishes for odd $a_\perp$. After the $x^\perp$ integration, we are left with a 2d integral of the form
\benn
\int dx^+ dx^- \frac{e^{iP\cdot x} }{ (x^+)^{A-a_+-\frac{a_\perp+1}{2}}  (x^-)^{A-a_--\frac{a_\perp+1}{2}} },
\eenn
which we already evaluated in section~\ref{sec:2Dformulas}. The final Fourier transform formula is thus\footnote{We have checked that this formula agrees with equation (2.12) of \cite{Gillioz:2018mto} for several examples.}
\be
\boxed{
\ba
\Ical_\bA(P) &\equiv e^{-\fr{i\pi}{2}(2A - a_+ - a_-)} \int d^3x \, e^{iP\cdot x}\, \frac{(x^+)^{a_+}(x^-)^{a_-}(x^\perp)^{a_\perp}}{(x^2)^A} \\
&= \frac{(1+(-1)^{a_\perp})\pi^2 \Gamma(\frac{a_\perp+1}{2}) \Gamma(A-\frac{a_\perp+1}{2}) P_+^{A-a_+-\frac{a_\perp+3}{2}} P_-^{A-a_--\frac{a_\perp+3}{2}} }{2^{A-\frac{a_\perp+3}{2}} \Gamma(A) \Gamma(A-a_+-\frac{a_\perp+1}{2}) \Gamma(A-a_--\frac{a_\perp+1}{2}) } \Theta(P_+)\Theta(P_-) \\
& \hspace{4.5in} (P_\perp = 0). 
\ea
}
\label{eq:IFinal}
\ee

\subsection{Three-point functions}
\label{subsec:3D3PF}

Similar to two-point functions, in 3d Lorentzian CFT three-point functions can be expanded into terms of the form
\be
\ba
&\langle \Ocal(x_1) \Ocal_R(0) \Ocal^\prime(x_3) \rangle \\
& \qquad \; \supseteq e^{i\theta_{\bA\bB\bC}} \frac{(x_1^+)^{a_+} (x_1^-)^{a_-} (x_1^\perp)^{a_\perp} (-x_3^+)^{b_+} (-x_3^-)^{b_-} (-x_3^\perp)^{b_\perp} (x_{13}^+)^{c_+} (x_{13}^-)^{c_-} (x_{13}^\perp)^{c_\perp}}{(x_1^2)^A (x_3^2)^B (x_{13}^2)^C},
\ea
\ee
where the overall phase is given by
\be
\theta_{\bA\bB\bC} \equiv \fr{\pi}{2}(a_+ + a_- + b_+ + b_- + c_+ + c_-) - \pi(A+B+C),
\ee
and for notational simplicity we have suppressed the factors of $i\epsilon$, which have the same structure as eq.~\eqref{eq:3d2ptWightman}. To obtain the contribution of these terms to the momentum space Wightman function, we need to evaluate the integral
\be
\Mcal_{\bA\bB\bC} (P,P^\prime) \equiv e^{i\theta_{\bA\bB\bC}} \int d^3x_1 \, d^3x_3 \, e^{i(P\cdot x_1 - P^\prime\cdot x_3)} \frac{\prod_{\mu=\pm,\perp} (x_1^\mu)^{a_\mu} (-x_3^\mu)^{b_\mu} (x_{13}^\mu)^{c_\mu}}{\left(x_1^2\right)^A \left(x_3^2\right)^B \left(x_{13}^2\right)^C}.
\label{eq:JDef}
\ee
Matrix elements of (general-spin) primary operators will be linear combinations of the functions $\Mcal_{\bA\bB\bC} (P,P^\prime)$,
\be
\langle \Ocal(P) | V | \Ocal^\prime(P^\prime ) \rangle = (2\pi)^2 \delta^2(\vec{P}-\vec{P}^\prime) \sum_{\bA\bB\bC} T_{\Ocal\Ocal^\prime\Ocal_R}^{\bA\bB\bC} \, \Mcal_{\bA\bB\bC} (P,P^\prime), 
\ee
for some coefficients $T_{\Ocal\Ocal^\prime\Ocal_R}^{\bA\bB\bC} $. Note that the delta function out front sets $\vec{P}=\vec{P}^\prime$. We will therefore impose this in evaluating $\Mcal_{\bA\bB\bC} (P,P^\prime)$. 

Ultimately, we will write down an explicit formula for $\Mcal_{\bA\bB\bC}$ in the simplified case when $a_- \!=\! b_- \!=\! c_- \!=\! 0$. This case is of particular interest, because it is sufficient for computing conformal truncation matrix elements in free scalar field theory, which is the immediate goal of this work. While the more general case of nonzero $a_-$, $b_-$, or $c_-$ is more complicated, we will explain how those formulas can also be computed. For now, we assume general arguments and will indicate when we choose to set the exponents $a_- \!=\! b_- \!=\! c_- \!=\! 0$. 

Our first step in evaluating $\Mcal_{\bA\bB\bC}$ is to rewrite the denominator in (\ref{eq:JDef}). We do this using the Anti-de Sitter (AdS) bulk representation for scalar three-point functions. Specifically, in $d$ spacetime dimensions, one has the formula~\cite{Freedman:1998tz}\footnote{Note that we have analytically continued the Euclidean expression in~\cite{Freedman:1998tz} to Lorentzian signature.}
\be
\ba
&\frac{e^{i\pi(A+B+C-\frac{1}{2})}}{(x_{1}^2)^A (x_{3}^2)^B (x_{13}^2)^C} \\
&= F^{(d)}(A,B,C) \! \int \! \frac{d^dx \, dw}{w^{d+1}} \! \left( \! \frac{w}{(x-x_1)^2-w^2} \! \right)^{A+C} \! \left( \! \frac{w}{x^2-w^2} \! \right)^{A+B} \! \left( \! \frac{w}{(x-x_3)^2-w^2} \! \right)^{B+C} \!,
\ea
\label{eq:Bulk3PF}
\ee
where the prefactor is defined as
\be
F^{(d)}(A,B,C) \equiv \frac{2\, \Gamma(A+C)\,\Gamma(A+B)\,\Gamma(B+C)}{\pi^{\frac{d}{2}} \, \Gamma(A)\,\Gamma(B)\,\Gamma(C)\, \Gamma(A+B+C-{\textstyle\frac{d}{2}}) }.
\label{eq:F}
\ee
The precise $i\epsilon$ structure of eq.~\eqref{eq:Bulk3PF} is discussed in section~\ref{sec:AdS}. The right-hand side represents the computation of a three-point Witten diagram in AdS$_{d+1}$ with bulk coordinates $(x^\mu,w)$. The utility of rewriting the denominator in this way is that the Fourier transform factorizes, at the cost of the extra integration variables. 

To proceed, we insert the bulk representation (\ref{eq:Bulk3PF}) into the expression for $\Mcal_{\bA\bB\bC}$ in (\ref{eq:JDef}). Then, in a slight abuse of notation, we relabel $x\rightarrow -x_2$, which will make the resulting expression somewhat more intuitive, even though this coordinate is really associated with the bulk AdS interaction point. Next, we can factorize the expression by shifting $x_{1,3} \rightarrow x_{1,3} - x_2$. The resulting expression for $\Mcal_{\bA\bB\bC}$ is 
\be
\ba
&\Mcal_{\bA\bB\bC}(P,P^\prime) \\
&\quad = F^{(3)}(A,B,C) \, e^{i\theta_{\bA\bB\bC}-i\pi(A+B+C-\frac{1}{2})} \! \int \! d^3x_1 \, d^3x_2 \, d^3x_3 \, e^{i(P\cdot x_1 - P^\prime\cdot x_3)} e^{-i(P_+ - P^\prime_+)x_2^+} \\
& \quad \times \int \! \fr{dw}{w^4} \! \left( \frac{w}{x_1^2-w^2} \right)^{A+C} \! \left( \frac{w}{x_2^2-w^2} \right)^{A+B} \! \left( \frac{w}{x_3^2-w^2} \right)^{B+C} \! \prod_{\mu=\pm,\perp} (x_{12}^\mu)^{a_\mu} (x_{23}^\mu)^{b_\mu} (x_{13}^\mu)^{c_\mu}.
\ea
\ee

As mentioned before, our strategy for evaluating this expression is to integrate over all of the transverse components first. With this in mind, we can expand out the $x^\perp_{ij}$ factors appearing in the numerator as 
\be
\ba
(x_{12}^\perp)^{a_\perp} (x_{23}^\perp)^{b_\perp} (x_{13}^\perp)^{c_\perp} &= \sum_{m_a=0}^{a_\perp} \sum_{m_b=0}^{b_\perp}  \sum_{m_c=0}^{c_\perp} (-1)^{b_\perp + m_a + m_c - m_b} {a_\perp \choose m_a} {b_\perp \choose m_b} {c_\perp \choose m_c} \\
& \qquad \times ( x_1^\perp )^{a_\perp + c_\perp -m_a - m_c}  ( x_2^\perp )^{m_a + m_b}  ( x_3^\perp )^{b_\perp - m_b + m_c}.
\ea
\label{eq:PerpExpansion}
\ee
Each term in this expansion gives rise to three factorized $x_i^\perp$ integrals of the same form as eq.~\eqref{eq:PerpIntegral}. Because the power of $x_i^\perp$ in the numerator of each of these integrals must be even for the expression to be nonzero, we therefore obtain the requirements
\be
\begin{rcases}
  & a_\perp + b_\perp + c_\perp \\
  & m_a + m_b \\
  & b_\perp - m_b + m_c
\end{rcases}
= \textrm{even.}
\ee

After expanding with \eqref{eq:PerpExpansion} and then evaluating the transverse integrals with \eqref{eq:PerpIntegral}, the resulting expression for $\Mcal_{\bA\bB\bC}$ contains sums over $m_{a,b,c}$ along with integrals of the general form
\be
\ba
&\int d^2x_1 \, d^2x_2 \, d^2x_3 \, e^{i(P\cdot x_1 - P^\prime\cdot x_3)} e^{-i(P_+ - P^\prime_+)x_2^+} \prod_{\mu=\pm} \left(x_{12}^\mu \right)^{a_\mu} \left(x_{23}^\mu \right)^{b_\mu} \left(x_{13}^\mu \right)^{c_\mu} \\
& \qquad \times \int \fr{dw}{w^4} \left( \frac{w}{2x_1^+ x_1^- - w^2} \right)^{A+C-\frac{a_\perp+c_\perp-m_a-m_c+1}{2}}  \left( \frac{w}{2x_2^+ x_2^- - w^2} \right)^{A+B-\frac{m_a+m_b+1}{2}} \\
& \qquad \times  \left( \frac{w}{2x_3^+ x_3^- - w^2} \right)^{B+C-\frac{b_\perp-m_b+m_c+1}{2}} w^{\frac{a_\perp+b_\perp+c_\perp+3}{2}},
\ea
\label{eq:JInt1}
\ee
where $d^2x \equiv dx^+ dx^-$. 

The next step is to integrate over $x_2^-$. \emph{Now} we assume that $a_- \!=\! b_- \!=\! c_- \!=\! 0$. For this case, the integral we need is worked out in appendix \ref{sec:useful}. We reproduce it here for the reader's convenience,
\be
\int_{-\infty}^\infty dx^- \frac{1}{(2x^+ x^- - w^2 - i\epsilon)^\De} = e^{i\pi(\De-\fr{1}{2})} \fr{\pi w^{2(1-\De)}}{\De-1} \de(x^+).
\label{eq:x2Int}
\ee
Since the $x_2^-$ integration is proportional to $\delta(x_2^+)$, we can trivially evaluate the integral over $x_2^+$, which simply sets $x_2^+=0$. These steps reduce the integral in (\ref{eq:JInt1}) to   
\be
\ba
&\int \frac{dw}{w^{A+B-\frac{1}{2}(a_\perp+b_\perp+c_\perp+m_a+m_b)}} \int d^2x_1 \, d^2x_3 \, e^{i(P\cdot x_1 - P^\prime\cdot x_3)}  \left( x_1^+ \right)^{a_+} \left( -x_3^+ \right)^{b_+}  \left( x_{13}^+ \right)^{c_+}   \\
&\qquad \times \left( \frac{w}{2x_1^+ x_1^- - w^2} \right)^{A+C-\frac{a_\perp+c_\perp-m_a-m_c+1}{2}} \left( \frac{w}{2x_3^+ x_3^- - w^2} \right)^{B+C-\frac{b_\perp-m_b+m_c+1}{2}}.
\ea
\label{eq:JInt2}
\ee

Before continuing, let us pause to make two comments. The first is that formula~(\ref{eq:x2Int}) assumes $\Delta >1$, or more generally $\Delta > \fr{d}{2}$ in higher dimensions. For conformal truncation matrix elements, this directly corresponds to assuming $\Delta_R>\fr{d}{2}$, where $\Delta_R$ is the dimension of the relevant operator perturbing the UV CFT. This is an important caveat for our formulas (as we alluded to earlier below (\ref{eq:2dME})). A physical understanding of this restriction comes from interpreting matrix elements as AdS Witten diagrams. We will discuss this in section~\ref{sec:AdS}.

Second, we comment on the case of nonvanishing $a_-$, $b_-$, or $c_-$. In that case, factors of $x^-$ would appear in the numerator on the left-hand side of \eqref{eq:x2Int}. As a result, the right-hand side would be modified to a derivative of a delta function, $\delta^{(n)}(x^+)$. Then, one would need to integrate by parts inside the $x_2^+$ integral before setting $x_2^+=0$. This can certainly be done, and the resulting integrals can be evaluated using the same steps discussed below. For our purposes, though, we are content to ignore these complications and focus on the case $a_- \!=\! b_- \!=\! c_- \!=\! 0$. 

Returning to the integral \eqref{eq:JInt2}, we can make progress with the following crucial observation: this integral is precisely the Fourier transform of a three-point function in 2d, written in a bulk representation! To see this, define
\be
\ba
A^\prime &\equiv A - \frac{a_\perp + c_\perp + m_b - m_c +1}{2}, \\
B^\prime &\equiv B - \frac{b_\perp + m_a + m_c + 1}{2}, \\
C^\prime &\equiv C + \frac{m_a + m_b}{2}.
\ea
\ee
Then the integral in \eqref{eq:JInt2} becomes
\be
\ba
& \int d^2x_1 \, d^2x_3 \, e^{i(P\cdot x_1 - P^\prime\cdot x_3)}  (x_1^+)^{a_+} (-x_3^+)^{b_+} (x_{13}^+)^{c_+} \\
& \quad \times \int \frac{dw}{w^{A^\prime + B^\prime + 1}} \left( \frac{w}{2x_1^+ x_1^- - w^2} \right)^{A^\prime + C^\prime} \left( \frac{w}{2x_3^+ x_3^- - w^2} \right)^{B^\prime + C^\prime} \\
& \qquad = \frac{e^{i\pi C^\prime}(A^\prime + B^\prime-1)}{2^{A^\prime+B^\prime+C^\prime}\pi F^{(2)}(A^\prime, B^\prime, C^\prime)} \int dx_1^+ \, dx_3^+ \frac{e^{i(P_+ x_1^+ - P^\prime_+ x_3^+)}}{(x_1^+)^{A^\prime-a_+} (-x_3^+)^{B^\prime-b_+} (x_{13}^+)^{C^\prime-c_+}} \\
& \qquad \quad \times \int dx_1^- \, dx_3^- \frac{e^{iP_- x_{13}^-}}{(x_1^-)^{A^\prime} (-x_3^-)^{B^\prime} (x_{13}^-)^{C^\prime}}.
\ea
\ee
We have therefore reduced our initial 3d formula to a sum of 2d integrals, which we already evaluated in the previous section. 

Putting all the pieces together leads us to our final expression for $\Mcal_{\bA\bB\bC}$ in the case of vanishing $a_-$, $b_-$, and $c_-$: 
\begin{equation}
\boxed{
\begin{aligned}
&\Mcal_{\bA\bB\bC}(P,P^\prime) \equiv e^{i\theta_{\bA\bB\bC}} \int d^3x_1 \, d^3x_3 \, e^{i(P\cdot x_1 - P^\prime\cdot x_3)} \frac{\prod_{\mu=+,\perp} (x_1^\mu)^{a_\mu} (-x_3^\mu)^{b_\mu} (x_{13}^\mu)^{c_\mu}}{(x_1^2)^A (x_3^2)^B (x_{13}^2)^C} \\
& = \frac{\pi^{7/2} \, 2^{5+\fr{a_\perp + b_\perp + c_\perp}{2}-A-B-C} P_-^{A + B + C - \fr{a_\perp + b_\perp + c_\perp}{2} - 3}}{\Gamma(A) \, \Gamma(B)\, \Gamma(C)\, \Gamma(A+B+C-{\textstyle\frac{3}{2}})} \, \Theta(P_-) \Theta(P_+) \Theta(P^\prime_+) \\
& \quad \times \sum_{m_a=0}^{a_\perp} \sum_{m_b=0}^{b_\perp} \sum_{m_c=0}^{c_\perp} (-1)^{m_a} {a_\perp\choose m_a}  {b_\perp\choose m_b}  {c_\perp\choose m_c} \\
& \quad \times {\textstyle\frac{1}{8}} \Big(1+(-1)^{a_\perp+b_\perp+c_\perp}\Big) \Big( 1+(-1)^{m_a+m_b} \Big) \Big( 1+(-1)^{b_\perp - m_b + m_c} \Big) \\
& \quad \times \frac{\Gamma({\textstyle\frac{a_\perp+c_\perp-m_a-m_c+1}{2}})\, \Gamma({\textstyle\frac{b_\perp - m_b +m_c +1}{2}})\, \Gamma({\textstyle\frac{2A+2B-m_a-m_b-3}{2}}) \,\Gamma({\textstyle\frac{2C+m_a+m_b}{2}}) \,\Gamma({\textstyle\frac{m_a+m_b+1}{2}}) }{\Gamma({\textstyle\frac{2B-2b_+-b_\perp - m_a - m_c-1}{2}}) \, \Gamma({\textstyle\frac{2A+2C-2a_+-2c_+-a_\perp-c_\perp+m_a+m_c-1}{2}}) } \\
& \quad \times P_+^{A + C - a_+ - c_+ - \fr{a_\perp + c_\perp - m_a - m_c + 3}{2}} P_+^{\prime B - b_+ - \fr{b_\perp + m_a + m_c + 3}{2}} \\
& \quad \times \phantom{}_2F_1\left({\textstyle\frac{2C-2c_++m_a+m_b}{2}}, {\textstyle\frac{-2B + 2b_+ +b_\perp + m_a + m_c +3}{2}}; {\textstyle\frac{2A + 2C - 2a_+ - 2c_+ -a_\perp - c_\perp + m_a + m_c - 1}{2}}; {\textstyle\frac{P_+}{P_+^\prime}}  \right) \\
& \hspace{2.5in} (\vec{P}=\vec{P}^\prime, \, P_\perp=0, \, P_+ < P_+^\prime, \, a_- \! = \! b_- \! = \! c_- \! = \! 0).
\end{aligned}
}
\label{eq:JFinal}
\end{equation}

%%%%%%%%%%%%%%%%%%%%%%%%%%%%%%%%%%%%%%%%%%%%%%%%%%%%%%%%%%%%%%%%%%%%%%%%%%%%%%%%%%%%%%%%%%%%%%%%%%%%
\section{AdS approach}
\label{sec:AdS}

So far, our approach has been to first compute the position space expression for CFT Wightman functions, then explicitly Fourier transform to obtain the corresponding momentum space result. However, one useful way to automatically enforce $d$-dimensional CFT kinematics is to work in $(d+1)$-dimensional Anti-de Sitter space, which effectively ``geometrizes'' the action of the conformal generators \cite{Sundrum:2011ic}. In this section, we reconsider matrix elements for scalar primary operators using AdS to illustrate this alternative approach. 

Because we are specifically interested in momentum space correlation functions, it is most natural to work in Poincar\'{e} patch coordinates $(x^\mu,w)$, where $x^\mu$ are the $d$-dimensional spacetime coordinates of the CFT, and $w$ is the bulk AdS coordinate (with the spatial boundary of AdS located at $w=0$). In these coordinates, the AdS metric is
\be
ds_{\textrm{AdS}}^2 = \fr{\eta_{\mu\nu} dx^\mu dx^\nu - dw^2}{w^2}.
\ee
Since two- and three-point functions are completely fixed (up to OPE coefficients) by conformal symmetry, they can always be rewritten as boundary correlation functions of an effective field theory in AdS, even for CFTs which have no large $N$ parameter or no clear holographic dual.

\begin{figure}[t!]
\begin{center}
\includegraphics[width=0.35\textwidth]{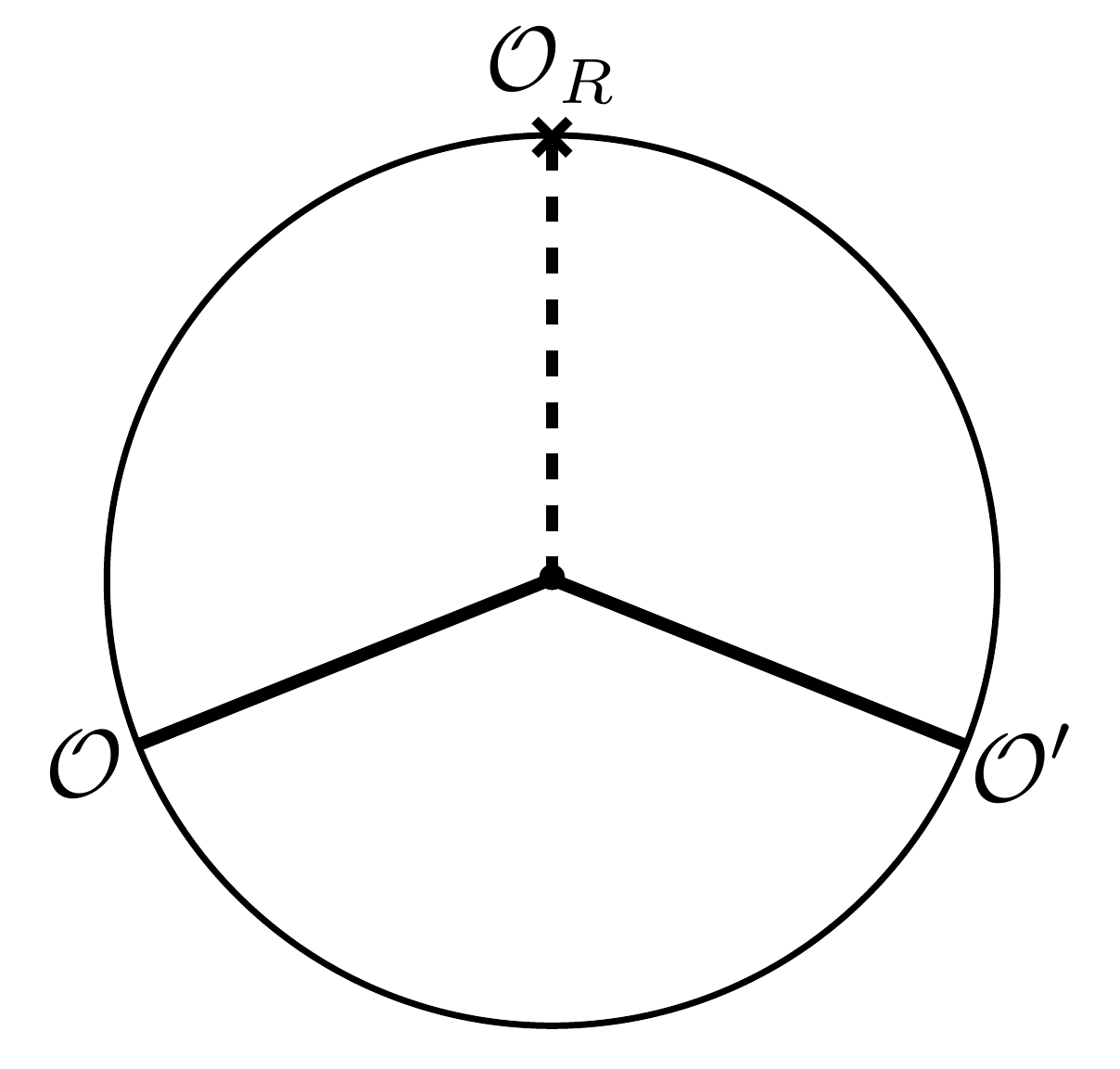}
\caption{Witten diagram for the CFT Wightman function $\<\Ocal\Ocal_R\Ocal'\>$. The solid lines for the two external operators represent Wightman bulk-to-boundary propagators, while the dashed line for the middle operator represents a time-ordered propagator.}
\label{fig:AdS3pt} 
\end{center}
\end{figure}

For example, consider a three-point Wightman function $\<\Ocal \Ocal_R \Ocal^\prime\>$ where all three operators are scalar primaries. As shown in figure~\ref{fig:AdS3pt}, this CFT correlator can be rewritten as an AdS Witten diagram containing a cubic interaction between three bulk scalar fields, whose masses $m^2$ are related to the scaling dimensions of these operators by
\be
m^2 = \De(\De-d).
\ee
This Witten diagram corresponds to a product of the three bulk-to-boundary propagators connecting the CFT operators to the bulk interaction point, with the location of this interaction integrated over all of AdS, resulting in the schematic expression
\be
\ba
&\<\Ocal(x_1) \Ocal_R(x_2) \Ocal^\prime(x_3)\> \\
&\qquad \quad \sim \int \! \fr{dw}{w^{d+1}} \! \int \! d^dx \, \<\Ocal(x_1) \phi(x,w)\> \<\mathcal{T}\{\Ocal_R(x_2) \phi_R(x,w)\}\> \<\phi^\prime(x,w) \Ocal^\prime(x_3)\>.
\ea
\ee

Note that, for this particular ordering of the CFT correlator, the bulk-to-boundary propagators associated with the external operators ($\Ocal$ and $\Ocal'$) are \emph{Wightman} propagators, which in position space take the form
\be
\<\phi(x^\mu,w) \Ocal(0)\> = \left( \fr{e^{-i\pi} w}{2(x^+ - i\epsilon)(x^- - i\epsilon) - |\vec{x}^\perp|^2 - w^2} \right)^\De,
\ee
where we've normalized this AdS propagator such that it reduces to a standard CFT two-point function in the limit $w \ra 0$,
\be
\<\Ocal(x)\Ocal(0)\> = \lim_{w\ra0} w^{-\De} \<\phi(x^\mu,w) \Ocal(0)\>.
\ee
The bulk-to-boundary propagator for the middle operator ($\Ocal_R$), however, is \emph{time-ordered},
\be
\<\mathcal{T}\{\phi(x^\mu,w) \Ocal(0)\}\> = \left( \fr{e^{-i\pi} w}{2 x^+ x^- - |\vec{x}^\perp|^2 - w^2 - i\epsilon} \right)^\De.
\ee

Combining these together, we thus obtain the full AdS representation for this scalar three-point function,
\be
\boxed{
\ba
&\<\Ocal(x_1) \Ocal_R(x_2) \Ocal^\prime(x_3)\> = C_{\Ocal\Ocal^\prime\Ocal_R} e^{-i\pi(\De+\De'+\De_R-\frac{1}{2})} F^{(d)}({\textstyle\frac{\De-\De^\prime+\De_R}{2}},{\textstyle\frac{\De^\prime-\De+\De_R}{2}},{\textstyle\frac{\De+\De^\prime-\De_R}{2}}) \\ 
& \qquad \times \int \fr{dw}{w^{d+1}} \int d^dx \left( \fr{w}{2(x^+ - x_1^+ + i\epsilon)(x^- - x_1^- + i\epsilon) - |\vec{x}^\perp - \vec{x}_1^\perp|^2 - w^2} \right)^\De \\
& \qquad \qquad \qquad \times \left( \fr{w}{2(x^+ - x_2^+)(x^- - x_2^-) - |\vec{x}^\perp - \vec{x}_2^\perp|^2 - w^2 - i\epsilon} \right)^{\De_R} \\
& \qquad \qquad \qquad \times \left( \fr{w}{2(x^+ - x_3^+ - i\epsilon)(x^- - x_3^- - i\epsilon) - |\vec{x}^\perp - \vec{x}_3^\perp|^2 - w^2} \right)^{\De^\prime},
\ea
}
\label{eq:AdS3pt}
\ee
where the prefactor $F^{(d)}(A,B,C)$ was defined previously in eq.~\eqref{eq:F}.

The advantage of this AdS representation is that it greatly simplifies the Fourier transform of the three-point function. Due to the factorized structure of~\eqref{eq:AdS3pt}, taking the Fourier transform with respect to the position $x_i$ simply recovers the corresponding momentum space bulk-to-boundary propagator. For the external operators $\Ocal$ and $\Ocal'$, we therefore obtain the momentum space Wightman propagator\footnote{See appendix~\ref{sec:useful} for a detailed derivation of the momentum space bulk-to-boundary propagators.}
\be
\ba
\int d^d x \, e^{i P\cdot x} \<\phi(x,w) \Ocal(0)\> &= \int d^d x \, e^{i P\cdot x} \left( \fr{e^{-i\pi} w}{2(x^+ - i\epsilon)(x^- - i\epsilon) - |\vec{x}^\perp|^2 - w^2} \right)^\De \\
&= \fr{\pi^{\fr{d}{2}+1}}{2^{\De-\fr{d}{2}-1} \G(\De)} \, \mu^{\De-\fr{d}{2}} w^{\fr{d}{2}} J_{\De-\fr{d}{2}}(\mu w) \, \Theta(\mu^2) \Theta(P_+),
\ea
\ee
where $J_\nu$ is a Bessel function of the first kind and $\mu^2 \equiv 2P_+ P_- - |\vec{P}_\perp|^2$. In the limit $w\ra0$, this propagator simply reduces to the Fourier transform of a CFT two-point Wightman function for a scalar primary,
\be
\lim_{w\ra0} w^{-\De} \left( \fr{\pi^{\fr{d}{2}+1}}{2^{\De-\fr{d}{2}-1} \G(\De)} \, \mu^{\De-\fr{d}{2}} w^{\fr{d}{2}} J_{\De-\fr{d}{2}}(\mu w) \right) = \fr{\pi^{\fr{d}{2}+1}\mu^{2\De-d}}{2^{2\De-d-1} \G(\De)\G(\De-\fr{d-2}{2})},
\ee
which agrees with our eqs.~\eqref{eq:2dInner} and \eqref{eq:IFinal} for the cases $d=2,3$.

For the middle operator $\Ocal_R$, the propagator is time-ordered, which has the corresponding momentum space expression
\be
\ba
\int d^d x \, e^{i P \cdot x} \<\mathcal{T}\{\phi_R(x,w) \Ocal_R(0)\}\> &= \int d^d x \, e^{i P \cdot x} \left( \fr{e^{-i\pi} w}{x^2 - w^2 - i\epsilon} \right)^{\De_R} \\
&= \fr{\pi^{\fr{d}{2}+1}}{2^{\De_R-\fr{d}{2}} \G(\De_R)} \mu^{\De_R-\fr{d}{2}} w^{\fr{d}{2}} H_{\De_R-\fr{d}{2}}(\mu w),
\ea
\label{eq:TOPropagator}
\ee
where $H_\nu$ is a Hankel function of the first kind. However, for computing Hamiltonian matrix elements in lightcone conformal truncation, we are specifically interested in the case where the spatial momenta $\vec{P} \equiv (P_-,\vec{P}_\perp)$ of the middle operator are set to zero, which is equivalent to taking $\mu^2 \ra 0$. In this limit, we find very different behavior for the Hankel function, depending on whether $\De_R$ is greater than or less than $\fr{d}{2}$:
\be
\lim_{\mu^2\ra0} H_{\De_R-\fr{d}{2}}(\mu w)\sim
\begin{cases}
\left(\fr{1}{\mu w}\right)^{\De_R-\fr{d}{2}} & (\De_R > {\textstyle\frac{d}{2}}), \\
\left(\fr{1}{\mu w}\right)^{\fr{d}{2}-\De_R} & (\De_R < {\textstyle\frac{d}{2}}).
\end{cases}
\ee

For $\De_R > \fr{d}{2}$, the powers of $\mu$ perfectly cancel, such that (\ref{eq:TOPropagator}) is finite for $\mu^2 \ra 0$. However, for $\De_R \leq \fr{d}{2}$, this limit is divergent. This is exactly the same behavior discussed in section~\ref{subsec:2D3PF}, where Hamiltonian matrix elements in 2d were shown to be divergent for $\De_R \leq 1$. From the AdS perspective, we thus see that this behavior is \emph{universal}, and arises specifically from the bulk-to-boundary propagator for $\Ocal_R$. We therefore generically expect these divergences to appear in \emph{any} matrix element with $\De_R \leq \fr{d}{2}$, regardless of the number of spacetime dimensions or the spins of the external operators $\Ocal$ and $\Ocal'$.\footnote{In principle, this divergence could be softened or removed if the bulk field $\phi_R$ is derivatively coupled, such that powers of momenta in the numerator cancel the overall factor of $1/\mu^{d-2\De_R}$, which can only happen if the external operators have nonzero spin. However, in $d=2$ we saw above that this does not happen when the external operators are primary, and instead requires a linear combination of primaries and descendants to successfully remove all divergences, which for free field theory corresponds to the Dirichlet basis discussed in~\cite{Katz:2016hxp}. It would be interesting to understand if this behavior generalizes in higher dimensions, or if there are particular polarization structures of primary operators which have finite matrix elements.}

For the finite case where $\De_R>\fr{d}{2}$, we can use this bulk-to-boundary propagator to compute the integral necessary for Hamiltonian matrix elements,
\be
\ba
\int d^{d-1} x \, \<\mathcal{T}\{\phi_R(x,w) \Ocal_R(0)\}\> &= \int d^{d-1} x \left( \fr{e^{-i\pi} w}{x^2 - w^2 - i\epsilon} \right)^{\De_R} \\
&= -\fr{i\pi^{\fr{d}{2}}\G(\De_R-\fr{d}{2})}{\G(\De_R)} \, w^{d-\De_R} \de(x^+) \qquad (\De_R > {\textstyle\frac{d}{2}}).
\ea
\ee

Using these expressions, we can now compute the Hamiltonian matrix element for three scalar primaries by taking the Fourier transform of eq.~\eqref{eq:AdS3pt}. The integrals over the three boundary positions $x_i$ simply convert the bulk-to-boundary propagators to momentum space, and the resulting integral for the interaction position $x$ trivially reproduces the overall momentum-conserving delta function $\de^{d-1}(\vec{P}-\vec{P}')$. We are therefore left with a single integral, over the bulk position $w$,
\be
\boxed{
\ba
&\int d^dx_1 \, d^{d-1}x_2 \, d^dx_3 \, e^{i(P\cdot x_1 - P^\prime\cdot x_3)} \, \<\Ocal(x_1) \Ocal_R(x_2) \Ocal^\prime(x_3)\> \\
& \quad = (2\pi)^{d-1} \de^{d-1}(\vec{P}-\vec{P}^\prime) \fr{C_{\Ocal\Ocal^\prime\Ocal_R} 2^{d+3-\De-\De^\prime} \pi^{d+2}\G(\De_R-\fr{d}{2})}{\G(\fr{\De+\De_R-\De^\prime}{2})\G(\fr{\De'+\De_R-\De}{2})\G(\fr{\De+\De^\prime-\De_R}{2})\G(\fr{\De+\De^\prime+\De_R-d}{2})} \\
& \quad \qquad \times \mu^{\De-\fr{d}{2}} \mu^{\prime \, \De^\prime-\fr{d}{2}} \Theta(\mu^2) \Theta(P_+) \Theta(\mu^{\prime \, 2}) \Theta(P^\prime_+) \int \fr{dw}{w^{\De_R-d+1}} \, J_{\De-\fr{d}{2}}(\mu w) \, J_{\De^\prime-\fr{d}{2}}(\mu^\prime w).
\ea
}
\label{eq:BulkIntegral}
\ee
Note that in obtaining this expression, we have only made two assumptions: all three operators are scalar primaries and $\De_R > \fr{d}{2}$. This expression therefore holds in \emph{any} number of dimensions and for \emph{any} choice of external momenta $P,P'$. Note that this integral representation exactly matches the $p_2 \ra 0$ limit of eq.~(3.30) in~\cite{Bautista:2019qxj}.

For the case where the external operators $\Ocal$ and $\Ocal'$ have nonzero spin, the resulting Fourier transform will have the same overall structure as eq.~\eqref{eq:BulkIntegral}. Specifically, the expression can be reduced to a single integral over the bulk point $w$, where the integrand is a product of the bulk-to-boundary propagators for the spinning AdS fields $\phi$ and $\phi'$, with the power of $w$ in the integration measure set by $\De_R$. The overall polarization structure, corresponding to a polynomial in $P$ and $P'$, is set by the particular choice of bulk interaction coupling the two spinning fields to the scalar field $\phi_R$.

For this simple case where all operators are scalars, we can evaluate the integral over $w$ to obtain the final Hamiltonian matrix element
\be
\boxed{
\ba
&\<\Ocal(P)|V|\Ocal^\prime(P^\prime)\> \\
& \qquad = (2\pi)^{d-1} \de^{d-1}(\vec{P}-\vec{P}^\prime) \fr{C_{\Ocal\Ocal^\prime\Ocal_R} 2^{2(d+1)-\De-\De^\prime-\De_R}\pi^{d+2}\G(\De_R-\fr{d}{2})}{\G(\fr{\De+\De_R-\De^\prime}{2})\G(\fr{\De'+\De_R-\De}{2})\G(\fr{\De+\De^\prime+\De_R-d}{2})}  \\
& \qquad \qquad \times \fr{\mu^{2\De-d} \mu^{\prime \, \De^\prime+\De_R-\De-d}}{\G(\fr{\De^\prime+\De_R-\De-d+2}{2})\G(\De-\fr{d-2}{2})} \Theta(\mu^2) \Theta(P_+) \Theta(\mu^{\prime \, 2}) \Theta(P^\prime_+) \\
& \qquad \qquad \times {}_2F_1\Big({\textstyle\frac{\De+\De^\prime-\De_R}{2}},{\textstyle\frac{\De-\De^\prime-\De_R+d}{2}};\De-{\textstyle\frac{d-2}{2}};{\textstyle\frac{\mu^2}{\mu^{\prime 2}}} \Big),
\ea
}
\label{eq:Bulk3pt}
\ee
where in evaluating this integral we've assumed (without loss of generality) that $\mu^{\prime 2} > \mu^2$. This expression agrees with eqs.~\eqref{eq:2dME} and \eqref{eq:JFinal} for $d=2,3$. In addition, this matrix element precisely matches the $p_0 \ra 0$ limit of the general momentum space three-point function in eq.~(41) of~\cite{Gillioz:2019lgs}.

%%%%%%%%%%%%%%%%%%%%%%%%%%%%%%%%%%%%%%%%%%%%%%%%%%%%%%%%%%%%%%%%%%%%%%%%%%%%%%%%%%%%%%%%%%%%%%%%%%%%
\section{Application: 3d $\phi^4$-theory}
\label{sec:FFT}

In this section, we use the Fourier transform formulas derived in section~\ref{sec:3Dformulas} to compute the conformal truncation matrix elements for 3d $\phi^4$-theory. The Lagrangian for this theory is
\be
\mathcal{L} = \frac{1}{2}\p_\mu\phi\p^\mu\phi - \fr{1}{2}m^2\phi^2 - \fr{1}{4!}\lambda\phi^4,
\ee
and the corresponding lightcone Hamiltonian is
\be
P_+ = \int d^2\vec{x} \left( \fr{1}{2}(\p_\perp\phi)^2 +  \fr{1}{2}m^2\phi^2 + \fr{1}{4!}\lambda\phi^4 \right).
\label{eq:P+Phi4}
\ee
The UV CFT is free, massless scalar field theory, which is then deformed by the relevant operators $\phi^2$ and $\phi^4$.

The conformal truncation basis is constructed using CFT operators. For a free massless scalar, these operators are built from derivatives acting on the field $\phi$. Due to the equation of motion $\p^2\phi=0$, we can choose to eliminate $\p_+$ derivatives in favor of working only with $\p_-$ and $\p_\perp$. Making this choice, we utilize the following notation to write down general operators. First, let $k$ be a two-component vector with a minus and transverse component,
\be
k= (k_-, k_\perp),
\ee
and for a single field $\phi$ define
\be
\p^k \phi \equiv \p_-^{k_-}\p_\perp^{k_\perp}\phi.
\ee
For an operator with $n$ fields, we use a vector $\bk$ 
\be
\bk = \left( k_1,\dots,k_n \right),
\ee
where again each $k_i= (k_{i-}, k_{i\perp})$, and define
\be
\p^{\bk} \phi \equiv \p^{k_1}\phi\cdots \p^{k_2}\phi.
\label{eq:MonomialDef}
\ee
We refer to $\p^{\bk} \phi$ as a \emph{monomial}. General operators are linear combinations of these monomials,
\be
\Ocal(x) = \sum_{\bk} C^\Ocal_{\bk} \, \p^{\bk}\phi(x).
\label{eq:OpDef}
\ee
Finally, it is useful to introduce shorthand notation for summations,
\be
|\bk_-| = \sum_i k_{i-}, \hspace{10mm} |\bk_\perp| = \sum_i k_{i\perp}, \hspace{10mm} |\bk| = |\bk_-| + |\bk_\perp|.
\ee

We will now use the formulas derived in section~\ref{sec:3Dformulas} to compute $\phi^4$-theory conformal truncation matrix elements for states created by individual monomials,
\be
|\p^{\bk}\phi(P)\> \equiv \int d^3x \, e^{-iP\cdot x} \p^{\bk}\phi(x)|0\>.
\ee
Since primary operators are linear combinations of monomials, the formulas below are sufficient to compute matrix elements for all operators. Without loss of generality, we work in the frame $P_\perp=0$, allowing us to apply the formulas of section~\ref{sec:3Dformulas} directly. We will only focus on the computation of matrix elements in this work, leaving the analysis of the resulting Hamiltonian and its eigenstates for the future.

It is worth commenting that we have independently checked the formulas below using Fock space methods, since our UV CFT is a free theory. Those methods are interesting in their own right and will be considered in more detail in an upcoming publication. In practice, however, we have found the Fourier transform formulas below to be much more efficient and amenable to numerical evaluation.

\subsection{Inner product}
\label{subsec:inner}

The inner product (\ref{eq:Inner}) between conformal truncation states is computed by a Fourier transform of a position-space two-point function. Our normalization of $\phi$ is such that its two-point function is 
\be
\langle \phi(x)\phi(0) \rangle = \frac{-i}{4\pi(x^2)^{\frac{1}{2}} }.
\ee
The two-point function of arbitrary monomials (\ref{eq:MonomialDef}) can be obtained via Wick contractions, with the result
\be
\ba
\langle \p^{\bk}\phi(x) \p^{\bkp}\phi(0) \rangle &= \frac{(-i)^n (-1)^{|\bk_-|+|\bkp_\perp|} (2x^+)^{|\bk_-|+|\bkp_-|} (2x^\perp)^{|\bk_\perp|+|\bkp_\perp|} }{ (4\pi)^n ( x^2 )^{|\bk|+|\bkp|+{\textstyle\frac{n}{2}}} } \\
& \qquad \times \sum_{M=0}^{{\textstyle\frac{1}{2}}\left( |\bk_\perp|+|\bkp_\perp|\right) } \left( \frac{x^2}{x^{\perp 2}} \right)^M W_{M,\bk,\bkp}.
\ea
\label{eq:Mon2PF}
\ee
In this formula, $n$ is the particle number (\emph{i.e.}~the number of $\phi$'s in each monomial), and we have defined the coefficient
\be
\ba
W_{M,\bk,\bkp} \equiv  \sum_{\text{Perms}(\bkp)} \sum_{\{ \bfm_\perp:\, |\bfm_\perp|=M, \, \bfm_\perp \leq \frac{1}{2}\left( \bk_\perp + \bkp_\perp \right)   \}} \prod_{i=1}^n \left( \frac{1}{2} \right)_{m_{i\perp}}  \left( \frac{1}{2} \right)_{ k_{i-} + k^\prime_{i-} }  & \\[5pt]
\times  {{k_{i\perp}+k^\prime_{i\perp}}\choose{2m_{i\perp}}}  \left( k_{i-} + k^\prime_{i-} + \frac{1}{2} \right)_{k_{i\perp} + k^\prime_{i\perp} -m_{i\perp}}.
\label{eq:A}
\ea
\ee
In this definition, the outer sum is over all permutations of $\bkp$, the inner sum is over all vectors $\bfm_\perp = (m_{1\perp},\dots,m_{n\perp})$ satisfying $|\bfm_\perp|=M$ and $m_{i\perp} \leq (k_{i\perp} + k^\prime_{i\perp})/2$ for each component, and $(x)_y = \Gamma(x+y)/\Gamma(x)$ are Pochhammer symbols. 

The expression in (\ref{eq:Mon2PF}) can be massaged in different ways using the fact that $x^2 = 2x^+ x^- - x^{\perp2}$. We have purposefully chosen an expression that avoids $x^-$ so that when we handle Hamiltonian matrix elements below, we will be precisely in the case $a_- = b_- = c_-=0$ of (\ref{eq:JFinal}) and will be able to apply that formula directly. 

With the two-point function (\ref{eq:Mon2PF}) in hand, we can directly apply the Fourier transform formula (\ref{eq:IFinal}) to obtain the conformal truncation inner product between monomials,
\begin{equation}
\boxed{
\begin{aligned}
& \langle \p^{\bk}\phi (P) | \p^{\bkp}\phi (P^\prime) \rangle =  (2\pi)^3 \delta^3(P-P^\prime)\, \frac{(-i)^{|\bk|}i^{|\bkp|}\pi^2 P_-^{|\bk_-|+|\bkp_-|} \mu^{|\bk_\perp|+|\bkp_\perp|+n-3} }{ (4\pi)^n \, 2^{n-4} \Gamma\left( \frac{|\bk_\perp|+|\bkp_\perp|+n-1}{2} \right) } \hspace{15mm} \\[5pt]
& \hspace{10mm} \times  \left( {\textstyle\frac{1\,+\,\left(-1\right)^{|\bk_\perp|+|\bkp_\perp|}}{2}} \right) \, \sum_{M=0}^{\frac{1}{2}\left( |\bk_\perp|+|\bkp_\perp|\right) } \frac{ \left(-1\right)^{\frac{|\bk_\perp|+|\bkp_\perp|}{2}-M} \Gamma\left( \frac{|\bk_\perp|+|\bkp_\perp|}{2} + \frac{1}{2} -M  \right) }{\Gamma\left( |\bk| + |\bkp| + \frac{n}{2} - M  \right) } W_{M,\bk,\bkp} \\
& \hspace{4.55in} (\mu^2 \equiv P^2, P_\perp=0).
\end{aligned}
}
\label{eq:InnerFinal}
\end{equation}

\subsection{$\phi^2$ interaction}
\label{subsec:phi2}

In this section, we will work out matrix elements of 
\be
V_{\phi^2} \equiv  \fr{1}{2} \int d^2 \vec{x}\, \phi^2,
\ee
\emph{i.e.}, the mass term in the lightcone Hamiltonian (\ref{eq:P+Phi4}). An important feature and simplification of lightcone quantization is that the vacuum is trivial~\cite{Leutwyler:1970wn,Maskawa:1975ky,Brodsky:1997de}. In particular, lightcone kinematics forbids particle creation from vacuum. Consequently, if $n$ is the number of incoming particles for the $\phi^2$ interaction, then the number of outgoing particles is also $n$. The processes $n\rightarrow n\pm2$, which would be present in equal-time quantization, vanish in lightcone quantization, providing a substantial simplification of the matrix elements.  

The operator $\phi^2$ happens to be very special for two reasons. First, $\Delta_{\phi^2} < \fr{d}{2}$ for $d<4$, which means that it can have divergent matrix elements, as was discussed in section~\ref{sec:AdS}. Second, its matrix elements are proportional to  $\delta(P_+-P_+^\prime)$, which resembles the kinematic structure of an inner product. Technically, in deriving the Fourier transform formula (\ref{eq:JFinal}), we assumed $\Delta_R >  \fr{d}{2}$ and $P_+ < P_+^\prime$, so naively this formula cannot be applied directly to $\phi^2$ matrix elements. 

Fortunately, these subtleties are easily evaded. First, we restrict our attention to external states with non-divergent matrix elements. These are precisely the ``Dirichlet" states of \cite{Katz:2016hxp}, which for free scalar field theory are linear combinations of monomials where each $k_{i-}\geq 1$. Second, we notice that $\phi^2$ matrix elements are nothing but sums over modified inner products, which trivializes their computation. This latter fact is easiest to see using Fock space language. We will cover Fock space methods in detail in a companion paper, but for now, we present the minimal ingredients needed to understand $\phi^2$ matrix elements. 

In lightcone quantization, the mode expansion for a free scalar $\phi$ in 3d is 
\be
\phi(x) = \int \frac{d^2 p}{ (2\pi)^2 \sqrt{2p_-} } \left( e^{-ip\cdot x} a_p + e^{ip\cdot x} a_p^\dagger     \right),
\label{eq:PhiMode}
\ee
where the creation and annihilation operators satisfy $[ a_p, a_q^\dagger ] = (2\pi)^2 \delta^2(p-q)$. Thus, the mode expansion of an $n$-particle conformal truncation state (\ref{eq:CTstate}) takes the form 
\be
|\mathcal{O}(P)\rangle = \frac{1}{n!} \int \left( \prod_{i=1}^{n} \frac{d^2 p_i}{ (2\pi)^2 \, 2p_{i-}}  \right) (2\pi)^3 \delta^3\left( \sum_{i=1}^n p_i - P \right) F_{\Ocal}(p) |p_1,\dots,p_n\rangle,
\ee
where 
\bea 
|p\rangle &\equiv& \sqrt{2p_-} a_p^\dagger |0\rangle \\[5pt]
F_{\Ocal}(p) &\equiv& \langle p_1,\dots,p_n | \Ocal(0) \rangle.
\eea
The inner product between two states takes the form
\be
\ba
& \langle \Ocal(P) | \Ocal^\prime(P^\prime) \rangle \\
& = (2\pi)^3 \delta^3(P-P^\prime) \frac{1}{n!} \int \left( \prod_{i=1}^{n} \frac{d^2 p_i}{ (2\pi)^2 2p_{i-}}  \right) (2\pi)^3 \delta^3\left( \sum_{i=1}^n p_i - P \right) F^*_{\Ocal}(p) F_{\Ocal^\prime}(p). 
\ea
\label{eq:FockInner}
\ee

The key observation, which is straightforward to check, is that matrix elements of $V_{\phi^2}$ are given by the same expression as the inner product (\ref{eq:FockInner}), \emph{except} for an extra factor of $\sum_{i=1}^n \frac{1}{p_{i-}}$ appearing in the integrand. We can interpret each $\fr{1}{p_{i-}}$ as the deletion of a minus derivative from $F_{\Ocal}$ (or equivalently, from $F_{\Ocal^\prime}$). An immediate consequence of this observation is that mass matrix elements are related to sums of inner products by the simple formula
\be
\boxed{
\langle \p^\bk\phi (P)| V_{\phi^2} | \p^{\bkp}\phi (P^\prime) \rangle = -i \sum_{i=1}^n \langle \p^{\bk^{(i)}} \phi(P) | \p^{\bkp}\phi (P^\prime) \rangle. 
}
\label{eq:MassFinal}
\ee

\noindent In this formula, inner products are given by (\ref{eq:InnerFinal}), where the vector $\bk^{(i)}$ is equal to $\bk$, \emph{except} that the single component $k_{i-}$ has been decremented by $1$ (which accounts for the presence of the factor $\frac{1}{p_{i-}}$ in the Fock space expression). This formula makes manifest the need to restrict to Dirichlet states, with $k_{i-}\geq 1$. This result trivializes mass matrix elements once inner products are known. We now turn to $\phi^4$ matrix elements.

\subsection{$\phi^4$ interaction}
\label{subsec:phi4}

In this section, we will work out matrix elements of 
\be
V_{\phi^4} \equiv  \fr{1}{4!} \int d^2 \vec{x}\, \phi^4,
\ee
\emph{i.e.}, the quartic term in the lightcone Hamiltonian (\ref{eq:P+Phi4}). As we have already mentioned, lightcone kinematics forbids particle creation from vacuum. Consequently, if $n$ is the number of incoming particles for the $\phi^4$ interaction, then the number of outgoing particles must be either $n$ or $n\pm 2$. The case $n\rightarrow n\pm4$, which would be present in equal-time quantization, vanishes in lightcone quantization. Additionally, $n\rightarrow n+2$ and $n+2\rightarrow n$ matrix elements are simply related by Hermitian conjugation. Hence, it is sufficient to just consider the possibilities $n\rightarrow n$ and $n\rightarrow n+2$ in turn. 

\subsubsection{$n\rightarrow n$}
\label{subsubsec:nn}

We start by assuming $n>2$. Then Wick contractions gives the position-space correlator
\be
\ba
\langle \p^{\bk}\phi(x_1) \[ \phi^4(x_2)/4! \] \p^{\bkp}\phi(x_3) \rangle = \sum_{k_{i,j}} \sum_{\kp_{r,s}} & \langle \p^{k_{i,j}}\phi(x_1) \[ \phi^4(x_2)/4! \] \p^{\kp_{r,s}}\phi(x_3) \rangle \\[5pt] 
\times & \langle \p^{\bk/k_{i,j}}\phi(x_1) \, \p^{\bkp/\kp_{r,s}}\phi(x_3) \rangle. 
\ea
\label{eq:NNWick}
\ee
In this equation, we are using the notation $k_{i,j} = (k_i,k_j)$, with $i<j$, and $\kp_{r,s}=(\kp_r, \kp_s)$, with $r<s$, to respectively denote the two incoming and two outgoing momenta that get contracted with $\phi^4$. The remaining $\bk/k_{i,j}$ incoming momenta and $\bk/\kp_{r,s}$ outgoing momenta are spectators that get contracted with each other. 

The correlator of the spectators is given by the general monomial two-point function that we computed in (\ref{eq:Mon2PF}). Meanwhile, the correlator containing the $2\rightarrow 2$ interaction with $\phi^4$ is given by
\be
\ba 
& \langle \p^{k_{i,j}}\phi(x_1) \[ \phi^4(x_2)/4! \] \p^{\kp_{r,s}}\phi(x_3) \rangle \\[5pt]
& = \frac{(-1)^{|k_{i,j-}|+|\kp_{r,s\perp}|} \, 2^{|k_{i,j}|+|\kp_{r,s}|}  }{(4\pi)^4} 
\sum_{m_i=0}^{\frac{1}{2}k_{i\perp}}  \sum_{m_j=0}^{\frac{1}{2}k_{j\perp}}  \sum_{m_r=0}^{\frac{1}{2}\kp_{r\perp}}  \sum_{m_s=0}^{\frac{1}{2}\kp_{s\perp}} \Omega^{k_i}_{m_i} \, \Omega^{k_j}_{m_j}  \, \Omega^{\kp_r}_{m_r}  \, \Omega^{\kp_s}_{m_s} \\[5pt]
& \times \frac{ \left( x_{12}^+ \right)^{|k_{i,j-}|} \left( x_{12}^\perp \right)^{|k_{i,j\perp}| - 2|m_{i,j}|}  \left( x_{23}^+ \right)^{|\kp_{r,s-}|} \left( x_{23}^\perp \right)^{|\kp_{r,s\perp}| - 2|m_{r,s}|}   }{ \left( x_{12}^2 \right)^{|k_{i,j}| - |m_{i,j}|+1}  \left( x_{23}^2 \right)^{|\kp_{r,s}| - |m_{r,s}|+1} },
\ea
\label{eq:2to2Correlator}
\ee
where we have defined
\be
\Omega^k_m \equiv  {{k_{\perp}}\choose{2m}} \left(\frac{1}{2} \right)_{k_{-}}  \left( \frac{1}{2} \right)_{m}  \left( k_{-} + \frac{1}{2} \right)_{k_{\perp}-m} 
\ee
and are using the summation notation
\be
|k_{i,j}| \equiv |k_i| + |k_j|, \hspace{5mm} |m_{i,j}| \equiv m_i + m_j,
\ee
with analogous notation for $\kp$ and also for minus or perpendicular components. 

Let $V_{\phi^4}^{(n\rightarrow n)}$ denote the $n\rightarrow n$ piece of $V_{\phi^4}$. We evaluate its matrix elements by substituting the appropriate expressions for the position-space correlators on the right-hand side of (\ref{eq:NNWick}) and then Fourier transforming using (\ref{eq:JFinal}). The result is
\begin{equation}
\boxed{
\begin{aligned}
& \langle \p^\bk\phi (P)| V_{\phi^4}^{(n\rightarrow n)} | \p^{\bkp}\phi (P^\prime) \rangle =  (2\pi)^2 \delta^2(\vec{P}-\vec{P}^\prime) \, \frac{(-i)^{|\bk|} i^{|\bkp|} 2^{|\bk|+|\bkp|}  }{ (4\pi)^{n+2} }     \\[5pt]
& \hspace{20mm} \times  \sum_{k_{i,j}} \sum_{\kp_{r,s}} \sum_{m_i=0}^{\frac{1}{2}k_{i\perp}}  \sum_{m_j=0}^{\frac{1}{2}k_{j\perp}}  \sum_{m_r=0}^{\frac{1}{2}\kp_{r\perp}}  \sum_{m_s=0}^{\frac{1}{2}\kp_{s\perp}} \sum_{M=0}^{\frac{1}{2}\left( |\bk_\perp| + |\bkp_\perp| - |k_{i,j\perp}| - |\kp_{r,s\perp}| \right)}    (-1)^{\fr{a_\perp+b_\perp+c_\perp}{2}} \,  \\[5pt]
& \hspace{20mm} \times \, \Omega^{k_i}_{m_i} \, \Omega^{k_j}_{m_j}  \, \Omega^{\kp_r}_{m_r}  \, \Omega^{\kp_s}_{m_s} \, A_{M,\bk/k_{i,j},\bkp/\kp_{r,s}}  \Mcal_{\bA\bB\bC}(P,P^\prime)  \\[15pt]
& \text{where}\hspace{3mm}  \bA = \left( |k_{i,j-}|,\,  0,\,  |k_{i,j\perp}|-2|m_{i,j}|,\,  |k_{i,j}|-|m_{i,j}|+1 \right) \\
& \phantom{\text{where}}\hspace{3mm} \bB  = \left( |\kp_{r,s-}|,\,  0,\,  |\kp_{r,s\perp}| - 2|m_{r,s}| ,\,  |\kp_{r,s}|-|m_{r,s}|+1 \right) \\
&  \phantom{\text{where}}\hspace{3mm} \bC  = \left( |\bk_-|+|\bkp_-|-|k_{i,j-}|-|\kp_{r,s-}|,\,  0,\,  |\bk_\perp|+|\bkp_\perp|-|k_{i,j\perp}|-|\kp_{r,s\perp}|-2M, \right. \\
& \hspace{22mm} \left. |\bk|+|\bkp|-|k_{i,j}|-|\kp_{r,s}|-M+n/2-1 \right) \\
& \hspace{95mm} (n>2, P_\perp =0, P_+ < P_+^\prime).
\end{aligned}
}
\label{eq:NNFinal}
\end{equation}
In this formula and the ones that follow, recall that bold symbols $\bA = (a_+,a_-,a_\perp,A)$ denote a collection of indices (see~(\ref{eq:bA})). 

The formula above is valid for $P_+ < P_+^\prime$, as this is the assumption that went into evaluating the Fourier transform $\Mcal_{\bA\bB\bC}(P,P^\prime)$ in~(\ref{eq:JFinal}). Equivalently, the formula above applies when the bra state has smaller invariant mass than the ket state, \emph{i.e.}, $P^2 < P^{\prime 2}$. Matrix elements for the opposite case $P_+ > P_+^\prime$ are easily obtained from~(\ref{eq:NNFinal}) via Hermitian conjugation. 

Note that we also assumed $n>2$ above. The case $n=2$ is simpler, because there are no spectators. The position-space correlator is simply (\ref{eq:2to2Correlator}), and the Fourier transform factorizes into a product of two-point function Fourier transforms, which can be done using our formula (\ref{eq:IFinal}). The final result for $2\rightarrow 2$ matrix elements is
\begin{equation}
\boxed{
\begin{aligned}
& \langle \p^\bk\phi(P) | V_{\phi^4}^{(2\rightarrow 2)} | \p^{\bkp}\phi,P^\prime \rangle =  (2\pi)^2 \delta^2(\vec{P}-\vec{P}^\prime) \, \frac{(-i)^{|\bk|} i^{|\bkp|} 2^{|\bk|+|\bkp|}  }{ (4\pi)^{4} }     \\[5pt]
& \hspace{20mm} \times \sum_{m_i=0}^{\frac{1}{2}k_{i\perp}}  \sum_{m_j=0}^{\frac{1}{2}k_{j\perp}}  \sum_{m_r=0}^{\frac{1}{2}\kp_{r\perp}}  \sum_{m_s=0}^{\frac{1}{2}\kp_{s\perp}}   (-1)^{\fr{a_\perp+a^\prime_\perp}{2}} \, \Omega^{k_i}_{m_i} \Omega^{k_j}_{m_j}  \, \Omega^{\kp_r}_{m_r}  \, \Omega^{\kp_s}_{m_s} \, 
\Ical_{\bA}(P) \, \Ical_{\bA^\prime}(P^\prime)
 \\[15pt]
& \text{where}\hspace{3mm}  \bA = \left( |\bk_-|,\, 0,\, |\bk_\perp|-2|m_{i,j}|,\, |\bk|-|m_{i,j}|+1 \right) \\
& \phantom{\text{where}}\hspace{3mm} \bA^\prime = \left( |\bkp_-|,\, 0,\, |\bkp_\perp|-2|m_{r,s}|,\, |\bkp|-|m_{r,s}|+1 \right) \\
& \hspace{87mm} (n=2, P_\perp=0, P_+ < P_+^\prime).
\end{aligned}
}
\label{eq:22Final}
\end{equation}

\subsubsection{$n\rightarrow n+2$}
\label{subsubsec:nn2}

The computation of $n\rightarrow n+2$ matrix elements proceeds just like the $n\rightarrow n$ case, with one exception. The $n$-particle state \emph{must} have invariant mass $P^{2}$ less than or equal to the $(n+2)$-particle invariant mass $P^{\prime 2}$. This means that matrix elements where $P_+ > P^\prime_+$ are just zero, which contrasts to the $n\rightarrow n$ case where $P_+ > P^\prime_+$ and $P_+ < P^\prime_+$ matrix elements are related by Hermitian conjugation. Otherwise, the computation proceeds as before. 

First, consider $n>1$. Then using Wick contractions, we can compute the $n\rightarrow n+2$ contribution to the $\phi^4$ three-point function as
\be
\ba
\langle \p^{\bk}\phi(x_1) \[ \phi^4(x_2)/4! \] \p^{\bkp}\phi(x_3) \rangle = \sum_{k_i} \sum_{\kp_{r,s,t}} & \langle \p^{k_{i}}\phi(x_1) \[ \phi^4(x_2)/4! \] \p^{\kp_{r,s,t}}\phi(x_3) \rangle \\[5pt] 
\times & \langle \p^{\bk/k_{i}}\phi(x_1) \, \p^{\bkp/\kp_{r,s,t}}\phi(x_3) \rangle. 
\ea
\label{eq:NN2Wick}
\ee
We are using the notation $k_i$ to denote the single incoming momentum and $\kp_{r,s,t}=(\kp_r, \kp_s,\kp_t)$, with $r<s<t$, to denote the three outgoing momenta that get contracted with $\phi^4$. The remaining $\bk/k_{i}$ incoming momenta and $\bk/\kp_{r,s,t}$ outgoing momenta are spectators that get contracted with each other. 

The correlator of the spectators was computed in (\ref{eq:Mon2PF}), and so the new ingredient needed is the correlator involving the $1\rightarrow 3$ interaction with $\phi^4$, which is given by 
\be
\ba 
& \langle \p^{k_{i}}\phi(x_1) \[ \phi^4(x_2)/4! \] \p^{\kp_{r,s,t}}\phi(x_3) \rangle \\[5pt]
& = \frac{(-1)^{|k_{i-}|+|\kp_{r,s,t\perp}|} \, 2^{|k_{i}|+|\kp_{r,s,t}|}  }{(4\pi)^4} 
\sum_{m_i=0}^{\frac{1}{2}k_{i\perp}}  \sum_{m_r=0}^{\frac{1}{2}\kp_{r\perp}}  \sum_{m_s=0}^{\frac{1}{2}\kp_{s\perp}}  \sum_{m_t=0}^{\frac{1}{2}\kp_{t\perp}}  \Omega^{k_i}_{m_i} \, \Omega^{\kp_r}_{m_r}  \, \Omega^{\kp_s}_{m_s} \, \Omega^{\kp_t}_{m_t} \\[5pt]
& \times \frac{ \left( x_{12}^+ \right)^{k_{i-}} \left( x_{12}^\perp \right)^{k_{i\perp} - 2m_{i}}  \left( x_{23}^+ \right)^{|\kp_{r,s,t-}|} \left( x_{23}^\perp \right)^{|\kp_{r,s,t\perp}| - 2|m_{r,s,t}|}   }{ \left( x_{12}^2 \right)^{|k_{i}| - m_{i}+\fr{1}{2}}  \left( x_{23}^2 \right)^{|\kp_{r,s,t}| - |m_{r,s,t}|+\fr{3}{2}} },
\ea
\label{eq:1to3Correlator}
\ee
where our summation notation is
\be
|\kp_{r,s,t}| \equiv |\kp_r| + |\kp_s| + |\kp_t|, \hspace{5mm} |m_{r,s,t}| \equiv m_r + m_s + m_t,
\ee
with analogous notation for minus and perpendicular components. 

Let $V_{\phi^4}^{(n\rightarrow n+2)}$ denote the $n\rightarrow n+2$ piece of $V_{\phi^4}$. Substituting our expressions for position-space correlators into the right-hand side of (\ref{eq:NN2Wick}) and Fourier transforming, we obtain the result
\begin{equation}
\boxed{
\begin{aligned}
& \langle \p^\bk\phi(P)| V_{\phi^4}^{(n\rightarrow n+2)} | \p^{\bkp}\phi(P^\prime) \rangle =  (2\pi)^2 \delta^2(\vec{P}-\vec{P}^\prime) \, \frac{(-i)^{|\bk|} i^{|\bkp|} 2^{|\bk|+|\bkp|}  }{ (4\pi)^{n+3} }     \\[5pt]
& \hspace{20mm} \times  \sum_{k_{i}} \sum_{\kp_{r,s,t}} \sum_{m_i=0}^{\frac{1}{2}k_{i\perp}}  \sum_{m_r=0}^{\frac{1}{2}\kp_{r\perp}}  \sum_{m_s=0}^{\frac{1}{2}\kp_{s\perp}} \sum_{m_t=0}^{\frac{1}{2}\kp_{t\perp}}  \sum_{M=0}^{\frac{1}{2}\left( |\bk_\perp| + |\bkp_\perp| - k_{i\perp} - |\kp_{r,s,t\perp}| \right)}    (-1)^{\fr{a_\perp+b_\perp+c_\perp}{2}} \,  \\[5pt]
& \hspace{20mm} \times \, \Omega^{k_i}_{m_i}  \, \Omega^{\kp_r}_{m_r}  \, \Omega^{\kp_s}_{m_s} \,  \Omega^{\kp_t}_{m_t} \, A_{M,\bk/k_{i},\bkp/\kp_{r,s,t}}  \Mcal_{\bA\bB\bC}(P,P^\prime)  \\[15pt]
& \text{where}\hspace{3mm}  \bA = \left( k_{i-}, \, 0, \, k_{i\perp}-2m_{i},\, |k_{i}|-m_{i}+1/2 \right) \\
& \phantom{\text{where}}\hspace{3mm} \bB = \left( |\kp_{r,s,t-}|,\, 0,\, |\kp_{r,s,t\perp}| - 2|m_{r,s,t}| ,\, |\kp_{r,s,t}|-|m_{r,s,t}|+3/2 \right) \\
&  \phantom{\text{where}}\hspace{3mm} \bC = \left( |\bk_-|+|\bkp_-|-k_{i-}-|\kp_{r,s,t-}|,\, 0,\, |\bk_\perp|+|\bkp_\perp|-k_{i\perp}-|\kp_{r,s,t\perp}|-2M, \right. \\
& \hspace{22mm} \left. |\bk|+|\bkp|-|k_{i}|-|\kp_{r,s,t}|-M+n/2-1/2 \right) \\
& \hspace{90mm} (n>1, P_\perp=0, P_+ < P_+^\prime).
\end{aligned}
}
\label{eq:NN2Final}
\end{equation}

Finally, let us consider the case $n=1$. One just needs to be a little careful about the relationship between the $1$-particle Fock state $|P\rangle$ and the operator $\phi(x)$ . Using the mode expansion in (\ref{eq:PhiMode}), it is straightforward to check that  $|P\rangle$ is related to the conformal truncation state $|\p_-\phi(P)\rangle$ by 
\be
|\p_-\phi(P) \rangle = i \pi\delta(P_+) |P\rangle. 
\label{eq:1pState}
\ee
Thus, the position-space correlator we need to consider is~(\ref{eq:1to3Correlator}) where $\p^{k_i}\phi = \p_-\phi$. The Fourier transform of this correlator factorizes into a product of two Fourier transforms of two-point functions. One of the Fourier transforms yields a delta function $\delta(P_+)$ that precisely cancels the delta function in (\ref{eq:1pState}) that comes from replacing $|\p_-\phi(P) \rangle$ with $|P\rangle$. The final answer for the $1\rightarrow 3$ matrix element is 
\begin{equation}
\boxed{
\begin{aligned}
& \langle P| V_{\phi^4}^{(1\rightarrow 3)} | \p^{\bkp}\phi(P^\prime) \rangle =  (2\pi)^2 \delta^2(\vec{P}-\vec{P}^\prime) \, \frac{i^{|\bkp|} 2^{|\bkp|}  }{ (4\pi)^{3} }     \\[5pt]
& \hspace{20mm} \times  \sum_{m_r=0}^{\frac{1}{2}\kp_{r\perp}}  \sum_{m_s=0}^{\frac{1}{2}\kp_{s\perp}} \sum_{m_t=0}^{\frac{1}{2}\kp_{t\perp}}  (-1)^{\fr{a^\prime_\perp}{2}} \,  \Omega^{\kp_r}_{m_r}  \, \Omega^{\kp_s}_{m_s} \,  \Omega^{\kp_t}_{m_t} \, \Ical_{\bA^\prime}(P^\prime) \\[15pt]
& \text{where}\hspace{3mm}  \bA^\prime = \left( |\bkp_-|,\, 0,\, |\bkp_{\perp}|-2|m_{r,s,t}|,\, |\bkp|-|m_{r,s,t}|+3/2 \right) \\
& \hspace{72mm} (n=1, P_\perp=0).
\end{aligned}
}
\label{eq:13Final}
\end{equation}

This concludes our computation of $\phi^4$-theory conformal truncation matrix elements. Though the main results we have derived look complicated, we emphasize that they are simple and computationally inexpensive to implement in, \textit{e.g.}, Mathematica. While they involve several nested sums, the summands are often simple rational functions with repeated arguments that can be memoized for rapid numerical evaluation. In short, the boxed formulas above are the basic ingredients needed to begin full-fledged numerical studies of $\phi^4$-theory in 3d.

In an upcoming publication~\cite{Upcoming}, we will use the above formulas to construct the Hamiltonian of $\phi^4$-theory and study its nonperturbative dynamics. To recap, in the framework of conformal truncation one views $\phi^4$-theory as free massless scalar field theory in the UV deformed by a $\phi^2$ mass term and a $\phi^4$ interaction. The formulas derived above can be used to construct the Hamiltonian in the following way. First, one constructs a basis of operators in free massless scalar field theory by writing general operators as linear combinations of monomials (see~\eqref{eq:OpDef}) and then performing Gram-Schmidt with respect to the inner product formula in~\eqref{eq:InnerFinal}. This yields a map between orthogonalized basis states and monomials. The $\phi^2$ and $\phi^4$ formulas presented above represent Hamiltonian matrix elements in ‘monomial’ space, which need to be rotated to ‘basis’ space using this map in order to obtain the final Hamiltonian. The technical details will be discussed in~\cite{Upcoming}. Finally, one numerically diagonalizes this Hamiltonian to study the resulting eigenvalues and eigenvectors and to compute other physical observables of interest. Since numerical diagonalization can be applied even at strong coupling, this procedure yields nonperturbative results for the dynamics of $\phi^4$-theory, using only momentum space correlators from free field theory.

%%%%%%%%%%%%%%%%%%%%%%%%%%%%%%%%%%%%%%%%%%%%%%%%%%%%%%%%%%%%%%%%%%%%%%%%%%%%%%%%%%%%%%%%%%%%%%%%%%%%
\section{Discussion and outlook}
\label{sec:conclusion}

To study any theory with conformal truncation, one needs to construct the Hamiltonian matrix elements for the relevant deformation(s) $\Ocal_R$, which can be written in the general schematic form
\be
\<\Ocal(P)|V|\Ocal'(P')\> = (2\pi)^{d-1} \de^{d-1}(\vec{P}-\vec{P}') \, C_{\Ocal\Ocal'\Ocal_R} \, \Mcal^{\Ocal_R}_{\Ocal\Ocal'}(P,P').
\ee
The Hamiltonian is therefore built from two key ingredients: the OPE coefficients $C_{\Ocal\Ocal'\Ocal_R}$ of the UV CFT and the kinematic function of external momenta $\Mcal^{\Ocal_R}_{\Ocal\Ocal'}(P,P')$, which is completely fixed by conformal symmetry and the operator dimensions and spins.

In this work, we have focused on the second ingredient, computing the universal kinematic functions for deformations of general CFTs in $1+1$ and $2+1$ dimensions. More specifically, we first computed the momentum space two-point Wightman functions, which are necessary for constructing orthonormal basis states, then we evaluated the Fourier transform of CFT three-point functions which correspond to Hamiltonian matrix elements in lightcone quantization. Given the spectrum of operators and OPE coefficients of a CFT, one can therefore use our results to construct the Hamiltonian for relevant deformations.

These results are particularly powerful in 2d, where there are an infinite number of minimal model CFTs whose spectrum and OPE coefficients are known. While particular deformations of these models are integrable, there remain many RG flows whose structure is still unexplored. So far, our conformal truncation method has only been used to study deformations of free theories, and 2d presents a rich arena in which to explore deformations of more general CFTs.

In 3d, we specifically computed the Fourier transform of individual ``monomials'' contributing to the full three-point function. While all Hamiltonian matrix elements can be expressed as a sum of these monomials, it would clearly be useful to organize our results into the particular linear combinations associated with primary operators, based on the spins of the two external operators. One particularly promising direction would be to generalize the AdS approach discussed in section~\ref{sec:AdS} to include operators with nonzero spin. More concretely, one would need to Fourier transform the spinning bulk-to-boundary propagators from~\cite{Costa:2014kfa} to momentum space, then decompose the set of bulk couplings between spinning AdS fields into the linearly independent tensor structures for CFT three-point functions. Some progress along these lines has been made in, \textit{e.g.}, \cite{Raju:2012zr,Raju:2012zs,Albayrak:2019asr,Albayrak:2019yve}. Towards this end, it may also be useful to impose conformal ward identities in momentum space~\cite{Gillioz:2019lgs} and/or use weight-shifting operators~\cite{Costa:2011dw,Karateev:2017jgd}. 

In this work, we have largely ignored the IR divergences that arise in matrix elements for operators with $\De_R \leq \frac{d}{2}$. These divergences are completely fixed by the kinematic structure of CFT three-point functions, and we should therefore be able to fully understand how they can be regulated or removed, so that we can study the finite eigenvalues of the Hamiltonian. In deformations of free field theories, these divergences reduce the CFT Hilbert space to the subspace of so-called ``Dirichlet'' states, which are particular linear combinations of primary operators with finite matrix elements~\cite{Katz:2016hxp}. However, it is currently unknown if this Dirichlet basis can be generalized to deformations of more general CFTs, or if a different approach must be taken instead.

One obvious future direction would be generalize our results to theories in $d \geq 4$. The main difference in higher dimensions is that the transverse direction is promoted to a vector. The simplest approach would be to represent $\vec{x}^\perp$ with spherical coordinates, in which case after evaluating the angular integrals the remaining expression should reduce to the form considered in section~\ref{sec:3Dformulas}.

Finally, it is worth noting that while in this work we have focused on the kinematic part of the Hamiltonian, we still lack efficient methods for computing the OPE coefficients, even in the simplest possible case of free field theory. The construction of a complete basis of primary operators in free theories and the computation of their OPE coefficients remains an (unappreciated) open problem, whose solution would be quite useful for a number of applications (see~\cite{Henning:2019mcv,Henning:2019enq} for recent progress). We encourage the community to consider this problem more thoroughly in the near future.

%%%%%%%%%%%%%%%%%%%%%%%%%%%%%%%%%%%%%%%%%%%%%%%%%%%%%%%%%%%%%%%%%%%%%%%%%%%%%%%%%%%%%%%%%%%%%%%%%%%%
\section*{Acknowledgments}

First we would like to thank Ami Katz for his continued advice and support throughout this project. We would also like to thank Simon Caron-Huot, Liam Fitzpatrick, Marc Gillioz, Brian Henning, Charles Hussong, Denis Karateev, and Jo\~{a}o Penedones for valuable discussions. We are grateful to the Abdus Salam International Centre for Theoretical Physics for hospitality while parts of this work were completed. This research was supported in part by Perimeter Institute for Theoretical Physics. Research at Perimeter Institute is supported by the Government of Canada through the Department of Innovation, Science and Economic Development and by the Province of Ontario through the Ministry of Research and Innovation. NA and ZK are supported by the Simons Collaboration on the Nonperturbative Bootstrap. ZK is also supported by the DARPA, YFA Grant D15AP00108. This research received funding from the Simons Foundation grant \#488649  (Simons Collaboration on the Nonperturbative Bootstrap). MW is partly supported by the National Centre of Competence in Research SwissMAP funded by the Swiss National Science Foundation.

\appendix
%%%%%%%%%%%%%%%%%%%%%%%%%%%%%%%%%%%%%%%%%%%%%%%%%%%%%%%%%%%%%%%%%%%%%%%%%%%%%%%%%%%%%%%%%%%%%%%%%%%%
\section{Summary of notation}
\label{sec:notation}

\subsection*{\underline{Fourier transforms}}
\be
\Ical_\bA(P) \equiv e^{-\fr{i\pi}{2}(2A - a_+ - a_-)} \int d^3x \, e^{iP\cdot x}\, \frac{(x^+)^{a_+}(x^-)^{a_-}(x^\perp)^{a_\perp}}{(x^2)^A}
\label{eq:app:I}
\ee

\be
\Mcal_{\bA\bB\bC} (P,P^\prime) \equiv e^{i\theta_{\bA\bB\bC}} \int d^3x_1 \, d^3x_3 \, e^{i(P\cdot x_1 - P^\prime\cdot x_3)} \frac{\prod_{\mu=\pm,\perp} (x_1^\mu)^{a_\mu} (-x_3^\mu)^{b_\mu} (x_{13}^\mu)^{c_\mu}}{\left(x_1^2\right)^A \left(x_3^2\right)^B \left(x_{13}^2\right)^C}
\label{eq:app:M}
\ee

\noindent In these formulas,
\begin{gather}
x^2 = 2(x^+ - i\epsilon)(x^- - i\epsilon) - x^{\perp 2} \\
\bA \equiv (a_+,a_-,a_\perp,A) \\
\theta_{\bA\bB\bC} \equiv \fr{\pi}{2}(a_+ + a_- + b_+ + b_- + c_+ + c_-) - \pi(A+B+C)
\end{gather}
Our results for $\Ical_\bA(P)$ and $\Mcal_{\bA\bB\bC} (P,P^\prime) $ are given in~(\ref{eq:IFinal}) and~(\ref{eq:JFinal}), respectively.

\vspace{5mm}
\subsection*{\underline{Monomial operators}}
Single $\phi$: 
\be
k= ( k_-, k_\perp ), \hspace{10mm} \p^k \phi \equiv \p_-^{k_-}\p_\perp^{k_\perp}\phi.
\ee
$n$ $\phi$'s:
\be
\bk = \left( k_1,\dots,k_n \right), \hspace{10mm} \p^{\bk} \phi \equiv \p^{k_1}\phi\cdots \p^{k_2}\phi.
\label{eq:app:bk}
\ee
Conformal truncation state:
\be
|\p^{\bk}\phi(P)\> \equiv \int d^3x \, e^{-iP\cdot x} \p^{\bk}\phi(x)|0\>.
\ee

\vspace{.5mm}
\subsection*{\underline{Summation}}
\vspace{-5mm}
\begin{gather}
|k_i| = k_{i-} + k_{i\perp} \\
|\bk_-| = \sum_{i=1}^n k_{i-}, \hspace{10mm} |\bk_\perp| = \sum_{i=1}^n k_{i\perp}, \hspace{10mm} |\bk| = \sum_{i=1}^n |k_i|
\end{gather}
Pair/triplet of $k$'s:
\begin{gather}
k_{i,j} \equiv ( k_i, k_j ), \hspace{10mm} k_{r,s,t} \equiv ( k_r, k_s, k_t ) \\[5pt]
|k_{i,j}| = |k_i| + |k_j|, \hspace{10mm} |k_{r,s,t}| = |k_r| + |k_s| + |k_t| 
\end{gather}
Pair/triplet of $m$'s:
\be
|m_{i,j}| = m_i + m_j, \hspace{10mm} |m_{r,s,t}| = m_r + m_s + m_t.
\ee

\vspace{1mm}
\subsection*{\underline{Useful coefficients}}
\be
\ba
W_{M,\bk,\bkp} \equiv  \sum_{\text{Perms}(\bkp)} \sum_{\{ \bfm_\perp:\, |\bfm_\perp|=M, \, \bfm_\perp \leq \frac{1}{2}\left( \bk_\perp + \bkp_\perp \right)   \}} \prod_{i=1}^n \left( \frac{1}{2} \right)_{m_{i\perp}}  \left( \frac{1}{2} \right)_{ k_{i-} + k^\prime_{i-} }  & \\[5pt]
\times  {{k_{i\perp}+k^\prime_{i\perp}}\choose{2m_{i\perp}}}  \left( k_{i-} + k^\prime_{i-} + \frac{1}{2} \right)_{k_{i\perp} + k^\prime_{i\perp} -m_{i\perp}} &
\ea
\label{eq:app:A}
\ee

\vspace{5mm}
\be
\Omega^k_m \equiv  {{k_{\perp}}\choose{2m}} \left(\frac{1}{2} \right)_{k_{-}}  \left( \frac{1}{2} \right)_{m}  \left( k_{-} + \frac{1}{2} \right)_{k_{\perp}-m} 
\ee
In~(\ref{eq:app:A}), the outer sum is over all permutations of $\bkp$ and the inner sum is over all vectors $\bfm_\perp = (m_{1\perp},\dots,m_{n\perp})$ satisfying $|\bfm_\perp|=M$ and $m_{i\perp} \leq (k_{i\perp} + k^\prime_{i\perp})/2$ for each component.

\section{Computation of AdS propagators}
\label{sec:useful}

In this appendix, we work out the following (closely-related) integral formulas
\bea
\mathrm{I}_1 &\equiv& \int d^dx \,e^{iP\cdot x} \left( \frac{w}{x^2 - w^2 - i\epsilon\, \mathrm{sgn}(t)} \right)^\Delta = \frac{e^{i\pi\Delta}(2\pi)^{\frac{d}{2}+1} \mu^{\Delta-\frac{d}{2}} w^{\frac{d}{2}} J_{\Delta-\frac{d}{2}}(\mu w) }{2^\Delta \Gamma(\Delta)}, \label{eq:useful1} \\[5pt]
\mathrm{I}_2 &\equiv& \int d^dx \,e^{iP\cdot x} \left( \frac{w}{x^2 - w^2 -i\epsilon} \right)^\Delta = \frac{e^{i\pi\Delta}(2\pi)^{\frac{d}{2}+1} \mu^{\Delta-\frac{d}{2}} w^{\frac{d}{2}} H_{\Delta-\frac{d}{2}}(\mu w) }{2^{\Delta+1} \Gamma(\Delta)}, \label{eq:useful2} \\[5pt]
\mathrm{I}_3 &\equiv& \int d^{d-1}\vec{x} \left( \frac{w}{x^2 - w^2 -i\epsilon} \right)^\Delta = \frac{e^{i\pi(\Delta-\fr{1}{2})}\pi^{\frac{d}{2}}\Gamma\left(\Delta-\frac{d}{2}\right) }{\Gamma(\Delta)} w^{d-\Delta} \delta\left(x^+ \right). \label{eq:useful3}
\eea
In these formulas, $J_\nu(x)$ is a Bessel function of the first kind, $H_\nu(x)$ is a Hankel function of the first kind, and we have introduced the notation $\mu^2 \equiv 2P_+ P_- - |\vec{P}_\perp|^2$. The first integral corresponds to the Fourier transform of a Wightman bulk-to-boundary propagator, while the second corresponds to a time-ordered bulk-to-boundary propagator. The third integral, which was used in the derivation of eq.~\eqref{eq:JFinal}, is the bulk-to-boundary propagator associated with an insertion of the Hamiltonian, which in lightcone quantization has $\mu^2 \ra 0$.

Before proceeding with the derivation of these formulas, it is worth pausing to comment on notation. Throughout this paper, we have been using lightcone coordinates $x=(x^+, \vec{x})$, where $x^+$ is lightcone time and $\vec{x}=(x^-, x^2,\dots,x^{d-1})$ are the spatial lightcone coordinates. However, in evaluating these integrals it will be somewhat simpler to work with the standard ``equal-time'' coordinates $x=(t,\vec{x}_\ET)$, where $\vec{x}_\ET = \left( x^1,\dots,x^{d-1}\right)$ are the usual spatial directions. We emphasize that $\vec{x}$ and $\vec{x}_\ET$ are different. 

With this notation in hand, let us begin with (\ref{eq:useful1}). We will work in the frame $P=(\mu,\vec{P}_\ET = 0)$. In this frame, 
\be
\textrm{I}_1 = w^\Delta \int d^{d-1}\vec{x}_\ET \int_{-\infty}^\infty dt \, f(t), 
\label{eq:I1Expr1}
\ee 
where
\be
f(t) = \frac{e^{i\mu t}}{\left(t+X-i\epsilon\right)^\Delta \left(t-X-i\epsilon\right)^\Delta},\hspace{10mm} X=\sqrt{|\vec{x}_\ET|^2 + w^2}.
\ee
In evaluating this integral we will assume that $\Delta$ is an integer, such that the singularities of $f(t)$ become poles, and analytically continue the resulting expression at the end. In this case, both poles of $f(t)$ are located above the real axis, such that this expression vanishes for $\mu < 0$. For positive $\mu$, we can close the contour in the upper half-plane and evaluate the residues for both poles, which for this particular function can be written in the convenient form
\be
\mathrm{Res}_{t=\pm X} [f(t)] = \frac{1}{2^{\Delta-1} \Gamma(\Delta)} \left( \frac{1}{X}\frac{\partial}{\partial X}  \right)^{\Delta-1} \left[ \pm \frac{e^{\pm i\mu X}}{2X} \right].
\label{eq:residue}
\ee
Summing both contributions, we obtain 
\be
\int_{-\infty}^{\infty} dt \, f(t) = 2\pi i \cdot  \frac{1}{2^{\Delta-1} \Gamma(\Delta)} \left( \frac{1}{X}\frac{\partial}{\partial X}  \right)^{\Delta-1} \left[ \frac{i\sin(\mu X)}{X} \right].
\ee

The residue formula above holds for arbitrary $X$. However, since in our particular case $X=\sqrt{|\vec{x}_\ET|^2 + w^2}$, we can further use the fact that 
\be
\frac{1}{X}\frac{\partial}{\partial X} = \frac{1}{w}\frac{\partial}{\partial w}.
\ee
This is useful, because we can pull the $w$-derivatives outside of the integral. Using these facts to perform the integral over $t$, we have 
\be
\ba
\mathrm{I}_1 &= -\frac{2\pi}{2^{\Delta-1}\Gamma(\Delta)} w^\Delta \left( \frac{1}{w}\frac{\partial}{\partial w} \right)^{\Delta-1} \int d^{d-1}\vec{x}_\ET \frac{\sin\left( \mu\sqrt{|\vec{x}_\ET|^2+w^2} \right)}{\sqrt{|\vec{x}_\ET|^2+w^2}} \\
&= -\frac{(2\pi)^{\frac{d}{2}+1}}{2^\Delta \Gamma(\Delta)} \mu^{1-\frac{d}{2}} w^\Delta  \left( \frac{1}{w}\frac{\partial}{\partial w} \right)^{\Delta-1} \left[  \frac{J_{1-\frac{d}{2}}(\mu w)}{ w^{1-\frac{d}{2}} } \right].
\ea
\ee
Finally, applying a derivative formula for Bessel functions,
\be
\left( \frac{1}{x}\frac{d}{dx}  \right)^m \left[ \frac{J_\alpha(x)}{x^\alpha}  \right] = (-1)^m \frac{J_{\alpha+m}(x)}{x^{\alpha+m}},
\ee
yields the final expression in (\ref{eq:useful1}).

The derivation of (\ref{eq:useful2}) is very similar. Again working in the frame $P=(\mu,\vec{P}_\ET = 0)$, the expression for $\mathrm{I}_2$ is identical to \eqref{eq:I1Expr1}, except that $f(t)$ is replaced by 
\be
g(t) = \frac{e^{i\mu t}}{\left(t+X+i\epsilon\right)^\Delta \left(t-X-i\epsilon\right)^\Delta},
\ee
\emph{i.e.}~one pole is above the real axis, and the other is below. The residues are still given by \eqref{eq:residue}, but this time only one of the two poles contributes, whether we close in the upper or lower half-plane. Following steps analogous to the ones described above, it is straightforward to check that for a single pole, instead of the Bessel function $J_\nu(x)$ in \eqref{eq:useful1}, one obtains the Hankel function $H_\nu(x)$ in \eqref{eq:useful2} (and an extra factor of $1/2$). 

Finally let us turn to \eqref{eq:useful3}. Heuristically, we expect the delta function $\delta(x^+)$ on the right-hand side for the following reason. Consider the $x^-$ integral. If $x^+\neq 0$, the integrand has a pole of order $\Delta$ in the $x^-$ complex plane, but the pole has vanishing residue (assuming $\Delta \neq 1$) such that $\mathrm{I}_3=0$. On the other hand, if $x^+=0$, the integrand is independent of $x^-$ and the integral diverges. These two observations suggest that the answer is proportional to $\delta(x^+)$. If we assume the presence of the delta function, the factor of $w^{d-\Delta}$ follows from dimensional analysis, and the constant prefactor can be computed by integrating both sides with respect to $x^+$. 

We can proceed a bit more systematically by introducing some infinitesimal lightcone momentum $P_-$, such that $P=(P_+, P_-, \vec{P}_\perp=0)$, and rewriting \eqref{eq:useful3} as 
\be
\mathrm{I}_3 = \lim_{P_-\rightarrow 0} \int \frac{dP_+}{2\pi} \, e^{-iP_+x^+} \int d^dx \, e^{iP\cdot x_1}  \left( \frac{w}{x^2 - w^2 -i\epsilon} \right)^\Delta
\ee
Note that we have added two integrations, over $P_+$ and $x^+$. The benefit of writing $\mathrm{I}_3$ this way is that we can use $\mathrm{I}_2$ to evaluate the integral over $x$, which gives
\be
\mathrm{I}_3 = \lim_{P_-\rightarrow 0} \int \frac{dP_+}{2\pi} \, e^{-iP_+x^+} \frac{e^{i\pi\Delta}(2\pi)^{\frac{d}{2}+1} \mu^{\Delta-\frac{d}{2}} w^{\frac{d}{2}} H_{\Delta-\frac{d}{2}}(\mu w)  }{2^{\Delta+1} \Gamma(\Delta)}.
\ee
Now we take the limit $P_-\rightarrow 0$, which corresponds to taking $\mu=\sqrt{2P_+P_-}\rightarrow 0$. Assuming $\nu = \Delta - \fr{d}{2} > 0$, for small arguments the Hankel function has the limit
\be
H_\nu(x) \rightarrow - \frac{i\Gamma(\nu)}{\pi} \left( \frac{2}{x} \right)^\nu + \dots \qquad (x \ra 0).
\ee
Plugging this in and simplifying, we find
\be
\mathrm{I}_3 = \frac{e^{i\pi(\Delta-\fr{1}{2})}\pi^{\frac{d}{2}} \Gamma\left(\Delta-\frac{d}{2}\right)}{\Gamma\left(\Delta\right)} w^{d-\Delta} \int \frac{dP_+}{2\pi} \, e^{-iP_+x^+},
\ee
which is precisely \eqref{eq:useful3} after evaluating the $P_+$ integral to obtain $\de(x^+)$.

\bibliographystyle{utphys}
\bibliography{Fourier}

\end{document}